\documentclass[12pt]{article}

\usepackage{ifpdf}
\usepackage{comment}
\usepackage{latexsym}
\usepackage{subfigure}
\usepackage{amssymb}
\usepackage{epsf}
\usepackage{cite}
\usepackage{amsmath}
\ifpdf
\usepackage{hyperref}
\else
\usepackage[hypertex]{hyperref}
\fi
\usepackage{graphicx}
\usepackage{setspace}
\usepackage{subfigure}
\usepackage{slashed}
\usepackage{cancel}
\usepackage{blkarray}

\makeatletter
\let\BA@quicktrue\BA@quickfalse
\makeatother

\makeatletter
    
    \@addtoreset{equation}{section}
  \makeatother

\newcommand{\lsim}{\lesssim}
\newcommand{\gsim}{\gtrsim}
\newcommand{\tr}{{\rm Tr}}

\begin{document}
\pagestyle{empty}

\begin{flushright}
\end{flushright}
    \vspace{2cm}
\begin{center}
{\bf\Large 
Natural Supersymmetry in Warped Space
}\\
    \vspace*{1.5cm}
{ Ben Heidenreich and Yuichiro Nakai} \\
    \vspace*{0.5cm}
{\it Department of Physics, Harvard University, Cambridge, MA 02138}\\
\end{center}
    \vspace*{1.0cm}

\begin{spacing}{1.2}
\begin{abstract}
\vspace{0.2cm}
{\normalsize
We explore the possibility of solving the hierarchy problem by combining the paradigms of supersymmetry and compositeness. Both paradigms are under pressure from the results of the Large Hadron Collider (LHC), and combining them allows both a higher confinement scale -- due to effective supersymmetry in the low energy theory -- and heavier superpartners -- due to the composite nature of the Higgs boson -- without sacrificing naturalness. The supersymmetric Randall-Sundrum model provides a concrete example where calculations are possible, and we pursue a realistic model in this context. With a few assumptions, we are led to a model with bulk fermions, a left-right gauge symmetry in the bulk, and supersymmetry breaking on the UV brane. The first two generations of squarks are decoupled, reducing LHC signatures but also leading to quadratic divergences at two loops. 
The model predicts light $W'$ and $Z'$ gauge bosons, and present LHC constraints on exotic gauge bosons imply a high confinement scale and mild tuning from the quadratic divergences, but the model is otherwise viable.
We also point out that R-parity violation can arise naturally in this context.
}
\end{abstract} 
\end{spacing}

\newpage
\baselineskip=18pt
\setcounter{page}{2}
\pagestyle{plain}
\baselineskip=18pt
\pagestyle{plain}

\setcounter{footnote}{0}

\section{Introduction} \label{sec:intro}


With the observation of a 125 GeV Higgs boson at the LHC~\cite{Higgs}, the final piece of the Standard Model (SM) is in place. However, the presence of a fundamental scalar combined with the lack of evidence for any new particles at the TeV scale presents a conundrum. The Standard Model must eventually incorporate gravity, hence it is at best an effective theory with a high cutoff scale. However, quantum corrections to the Higgs quadratic coupling are large, and scale quadratically with the cutoff, indicating a need for a fine-tuned cancellation between these corrections and the bare Higgs quadratic arising from a more fundamental theory. This is the hierarchy problem, which has motivated both theoretical and experimental study of beyond-the-Standard-Model (BSM) physics at or just above the electroweak symmetry breaking (EWSB) scale. While there is strong indirect evidence for BSM physics from cosmological considerations, such as the need for inflation, dark matter and successful baryogenesis, this physics is not yet associated to any particular energy scale, and need not involve new colored particles, making it easy to hide at the LHC. Therefore, there is as yet no evidence against a ``desert'' above the electroweak scale, with no new particles appearing at the LHC and a fine-tuned Higgs potential. The long-held assumption of electroweak naturalness is thus in question.

Two central ideas that have long played a role in solutions to the hierarchy problem are those of supersymmetry (SUSY) and compositeness. Supersymmetry solves the problem by cancelling the quadratic divergences between bosonic and fermionic loops, at the expense of requiring a superpartner of opposite statistics and like gauge quantum numbers for every particle in the Standard Model. By contrast, composite models postulate that the Higgs is a bound state of some new strongly-interacting dynamics, e.g.\ a pseudo-Goldstone boson of a QCD-like theory. 

These two approaches share some common problems as well as unique problems of their own. In their simplest forms, both predict large flavor-changing neutral currents (FCNCs) and rapid proton decay. Supersymmetry -- which predicts many new light particles -- struggles to accommodate ever more stringent LHC constraints. Moreover, the minimal supersymmetric model predicts a Higgs boson mass well below what is observed, unless loop corrections are employed to raise it at the expense of some fine tuning and a ``little hierarchy.'' By contrast, composite models are constrained by electroweak precision measurements, requiring a relatively high composite scale and fine tuning to achieve a light Higgs mass.

The prototypical example of a supersymmetric solution to the hierarchy problem is the Minimal Supersymmetric Standard Model (MSSM), which pairs every particle in the Standard Model with a superpartner and includes two Higgs doublets. The renormalizable interactions include both lepton- and baryon-number violating operators, so R-parity (under which all superpartners are odd) is usually imposed on the theory.
The SUSY-breaking mass terms and interactions generically induce large FCNCs, requiring some special flavor structure to be compatible with light superpartners, such as minimal flavor violation~\cite{D'Ambrosio:2002ex}. A similar organizing principle can also solve the proton decay problem 
without the imposition of R-parity~\cite{MFVSUSY}.
 
 The tree-level Higgs mass in the MSSM cannot exceed the $Z$ boson mass, and radiative corrections are needed, requiring a heavy stop and moderate tuning. An alternative is to introduce an additional singlet $S$ with a tree-level superpotential $S H_u H_d$, the next-to-minimal supersymmetric standard model (NMSSM)~\cite{NMSSM}. This increases the Higgs quartic coupling, allowing a heavier tree-level mass while also solving the $\mu$ problem. However, the required superpotential coupling is rather large, and may have a Landau pole if the cutoff of the theory is high.

Recent searches for supersymmetry at the LHC place strong constraints on the stop mass~\cite{stopbound}, suggesting some degree of tuning~\cite{naturalreview}. R-parity violation in the form of baryon number violation (BNV) can erase these constraints at the expense of removing a natural dark matter candidate (the lightest superpartner) from the theory, but equally stringent constraints on the gluino mass are virtually unaffected~\cite{Evans:2013jna,ATLAS-CONF-2013-091}, and the tuning problem persists.

Composite models come in many forms, but we focus on Randall-Sundrum (RS) models~\cite{Randall:1999ee} in this work. RS models are based on a warped extra dimension bounded by branes at either end. By the AdS/CFT correspondence~\cite{AdS/CFT,holography}, RS models are dual to (approximate) conformal field theories (CFTs), where the fifth dimension (in particular, the warp factor) is dual to the renormalization scale and the boundary branes are correspondingly labeled the infrared (IR) and ultraviolet (UV) branes. The appearance of an IR brane is dual to confinement in the four-dimensional theory, spontaneously breaking the approximate conformal invariance. By contrast, the UV brane is dual to an ultraviolet cutoff near the four-dimensional Planck scale.\footnote{Due to the presence of the UV brane the four-dimensional theory is gravitational, unlike in the usual AdS/CFT correspondence.} The AdS/CFT correspondence improves computability in both the four- and five-dimensional pictures, and we make repeated reference to it throughout our paper.

The RS solution to the hierarchy problem is to localize the Higgs field on the IR brane, corresponding to a composite Higgs in the CFT dual. Due to warping, the effective Planck scale on the IR brane is exponentially suppressed, cutting off five-dimensional loop corrections to the Higgs potential at a low scale. This scale, which we refer to as the (effective) compactification scale, is dual to the confinement scale in four dimensions, where loop corrections to the Higgs are cut off due to its composite nature.

The original RS model localizes all SM fields on the IR brane. However, this leads to severe flavor violation and proton decay problems due to dimension-six four-Fermi operators suppressed by the confinement scale. A simple solution to the flavor problem is to place the SM fermions and gauge bosons in the bulk~\cite{Grossman:1999ra,Chang:1999nh,Gherghetta:2000qt,Yukawa,Agashe:2004cp}. With appropriate boundary conditions, there is a chiral zero mode whose wavefunction profile depends exponentially on the mass term in the bulk.
Localizing the Higgs on the IR brane with anarchic order-one couplings to the bulk fermions, the profiles of the fermion zero modes can be adjusted to reproduce the observed Yukawa couplings in the low energy theory. Since the first- and second-generation fermions are localized towards the UV brane, they inherit substantial flavor protection from the ``RS-GIM mechanism.'' A mild flavor problem remains which can be addressed in various ways~\cite{flavor,Csaki:2008eh}. An added benefit of this approach is that it explains the hierarchical Yukawa couplings in terms of order-one differences in the bulk masses. In CFT language, the fundamental quarks couple to irrelevant CFT operators, so that their couplings to the CFT in the deep infrared are small and depend exponentially on the anomalous dimensions of these operators.

The proton decay problem is improved but not solved by the use of bulk fermions, but can be fixed by imposing a discrete symmetry and/or using horizontal symmetries.
However, the Higgs boson generically acquires a mass at the TeV scale, in tension with its observed 125 GeV mass.
Moreover, the sign of the Higgs quadratic term is not controlled and
the RS model does not provide a dynamical explanation for electroweak symmetry breaking.\footnote{A possible non-supersymmetric solution to these Higgs-sector problems is to realize the Higgs field as the fifth component of a bulk gauge field, dual to a pseudo Nambu-Goldstone-boson Higgs in CFT language~\cite{PNGH}. (See~\cite{SPNGH} for an attempt at a supersymmetric UV completion.) We do not consider this mechanism further in the present work.}

While some leeway still exists, it seems likely that neither supersymmetry nor compositeness can completely solve the hierarchy problem in their most minimal form. We are left with two alternatives. On the one hand, we could accept some yet-to-be-determined degree of fine tuning, abandoning or revising the naturalness paradigm. The problem with this approach is that -- without naturalness -- there is no good reason to expect BSM physics to be visible at the LHC. Even if new physics exists at the TeV scale, without a connection to naturalness too many possibilities exist for a targeted study, and there is no guarantee that it will be detected by the LHC. Anthropic arguments are not yet sufficiently refined to replace naturalness as a predictive framework, and the end result may be strong cosmological evidence for BSM physics without any indication of its type or energy scale.

The alternative is to pursue new approaches to the hierarchy problem, including less minimal realizations of supersymmetry and/or compositeness. Understanding this larger class of models is necessary to definitively establish whether the Higgs potential is fine-tuned, regardless of any theoretical prejudice for simpler models.\footnote{Indeed, string theory seems to exhibit a preference for non-minimal models.}

In this work, we consider a supersymmetric composite model. The combination of these two paradigms provides several benefits. By introducing a composite Higgs, we can eliminate fine tuning above the confinement scale, whereas effective supersymmetry~\cite{naturalSUSYspectrum} -- consisting of light Higgsinos, stops, and gauginos -- controls fine tuning below the confinement scale. The remaining squarks and sleptons can be heavy or even decoupled, relaxing LHC constraints on supersymmetry. The Higgs is naturally light due to effective supersymmetry, but the low cutoff allows us to introduce large tree-level couplings without fear of a Landau pole, raising the Higgs mass to its observed value. Due to effective supersymmetry, the confinement scale can be relatively large, evading electroweak precision measurements, flavor constraints, and direct searches for Kaluza-Klein (KK) modes. Finally, the underlying supersymmetry facilitates detailed computations of corrections to the Higgs potential,\footnote{However, decoupling the first two generations of squarks leads to hard SUSY breaking and greater computational difficulties.} allowing us to quantify naturalness.

A generic problem with non-minimal SM extensions is the need to introduce an additional energy scale besides the weak scale~\cite{Randall:2012dm}, leading to a coincidence problem. Composite SUSY models are no exception. In this case, the required coincidence is between the confinement scale and the soft SUSY-breaking scale (in particular, the scale of the gaugino masses). Heuristically, this suggests that SUSY breaking should trigger confinement in the strongly-coupled ultraviolet theory, but the details may not be so simple. While we recognize this to be an important problem, we will not address it explicitly in the present work.

To make progress, we consider the supersymmetric Randall-Sundrum model~\cite{Gherghetta:2000qt,SUSYRS,Marti:2001iw,Partly,Sundrum:2009gv,Gherghetta:2011wc,Larsen:2012rq} (cf.~\cite{CompositeSUSY}). While other choices are possible, we focus on a straightforward supersymmetrization of the ordinary RS model with the addition of a SUSY-breaking sector on the UV brane. The first two generations of squarks are localized towards the UV brane as required by the RS GIM mechanism, and couple directly to the SUSY breaking sector. Gauginos have a flat profile in the bulk, and also couple directly to SUSY breaking. However, the resulting gaugino masses are suppressed relative to the first two generations of squarks. In many cases, the suppression factor is large, and the first two generations of squarks decouple, with small gaugino masses receiving competitive contributions from anomaly and radion mediation. The Higgs and the stops are composite, with no direct coupling to the SUSY-breaking sector and soft masses generated by gaugino mediation and/or anomaly and radion mediation.

Models of this type have been considered previously in e.g.~\cite{Sundrum:2009gv,Gherghetta:2011wc}. The purpose of the present work is to pursue these ideas in an explicit model which is as realistic and concrete as possible. We encounter a number of issues which will affect earlier models as well, such as two-loop quadratic divergences due to decoupling the first two generations of squarks and light exotic gauge bosons necessitated by a previously recognized $U(1)$ $D$-term problem~\cite{Sundrum:2009gv,Strassler:2003ht}. These issues are related to the use of the RS GIM mechanism to solve the flavor problem, and may have analogues in other composite SUSY models. Nonetheless, we obtain a model that is viable with only minimal tuning.

Despite the protection offered by compositeness, the large masses of the decoupled squarks can still be problematic in one way. Generically, they generate a large hypercharge $D$-term, which induces a relevant deformation of the superconformal field theory (SCFT), spoiling the IR dynamics and introducing large fine tuning. To avoid this situation, the SM gauge group must be extended~\cite{Sundrum:2009gv}. We analyze the possible extensions in detail, and find that the only viable possibility is a left-right symmetric extension, e.g.\ the minimal left-right model, where other extensions lead to light charged and/or colored exotics excluded by the Large Electron-Positron Collider (LEP) and the LHC. Even the left-right extension leads to surprisingly light exotic gauge bosons, analogous to Higgsless models~\cite{Higgsless, Csaki:2005vy}. These $W'$ and $Z'$ gauge bosons are already strongly constrained by LHC results, requiring a relatively high confinement scale and mild fine tuning.

The SUSY RS model has a proton decay problem independent of whether R-parity is imposed or not~\cite{Gherghetta:2000qt}.\footnote{R-parity violation was originally discussed in~\cite{RPV}. See e.g.~\cite{Barbier:2004ez} for a review. Recent attempts at model building include~\cite{MFVSUSY,RPVmodel}.} To solve this problem, we impose a discrete lepton number symmetry, which stabilizes the proton regardless of whether R-parity is imposed. In particular, baryon number violation is allowed, and can lead to interesting phenomenology. The RS GIM mechanism is effective at suppressing BNV operators on the IR brane,\footnote{This is similar to MFV SUSY~\cite{MFVSUSY}, though the model is not strictly MFV.} whereas BNV operators on the UV brane are harmless once the first two generations of squarks are decoupled. We will show that introducing order-one anarchic couplings leads to a promptly decaying LSP, relaxing LHC constraints on missing energy while also satisfying low-energy constraints from $n - \bar{n}$ oscillations and dinucleon decay. This removes LHC constraints on the stop mass (see e.g.~\cite{Kats:2012ym}), reducing the tension between LHC results and naturalness. Nonetheless, our model is also viable with R-parity conservation.

The rest of the paper is organized as follows.
In section~\ref{sec:SUSYRS}, we review the SUSY RS model, placing
the standard model fields in the bulk of the warped extra dimension
and explaining the Yukawa hierarchies via their wavefunction profiles.
In section~\ref{sec:SUSYbreaking}, we discuss SUSY breaking in our scenario,
explaining possible sources of the gaugino and scalar soft masses.
The degree of fine tuning is then estimated.
In section~\ref{sec:Higgs}, we comment on the Higgs sector in our model, introducing
a singlet field to obtain a viable mass spectrum in the Higgs sector
as in the NMSSM.
The cosmological domain wall problem and the strong CP problem are also addressed.
In section~\ref{sec:RPV}, we introduce R-parity violating (RPV) couplings into the model and discuss experimental constraints
from low-energy measurements and collider experiments.
In section~\ref{sec:unification}, we extend the standard model gauge group to prevent a large hypercharge $D$-term and consequent fine tuning. We show that only one extension is viable, and discuss its low-energy signatures, some of which appear well below the IR scale.
In section~\ref{sec:flavor}, we comment on the constraints from flavor physics
and a possible flavor protection mechanism with $U(1)$ horizontal symmetries.
In section~\ref{sec:conclusions}, we conclude and discuss possible future directions.

\section{The SUSY RS model} \label{sec:SUSYRS}

In this section, we review the supersymmetric Randall-Sundrum (RS) model. 
We summarize the description of bulk vector and hypermultiplets coupled to IR-brane localized Higgs multiplets, and
 show that the Yukawa hierarchies are explained by the wavefunction profiles of the bulk matter fields.
These profiles will be important for estimating the size of RPV couplings, as discussed in section~\ref{sec:RPV}.

\subsection{Supersymmetric fields in the bulk}

We consider a 5D warped space with the extra dimension compactified on an $S^1/{\mathbb Z}_2$ orbifold:
$0 \leq |y| \leq \pi R$.
The spacetime metric is given by~\cite{Randall:1999ee}
\begin{equation}
ds^2 = e^{-2k |y|} \eta_{\mu\nu} dx^\mu dx^\nu + dy^2,
\end{equation}
where $\eta_{\mu\nu} = {\rm diag} (-1,1,1,1)$ and $k$ is the AdS curvature,
which is somewhat smaller than the 5D cutoff scale $\Lambda_5$.
The 4D Planck scale $M_4$ is related to the 5D Planck scale $M_5$ by $M_4^2 \simeq M_5^3 / k$.
Here, we consider a scenario where the (effective) compactification scale $k' \equiv k e^{-k \pi R}$ is near the TeV scale,\footnote{The mass of the lightest KK mode is of order $\pi k'$.}
which corresponds to $k R \sim 10$.
The UV (IR) brane is located at the orbifold fixed point, $y^\ast = 0$ ($y^\ast = \pi R$). 

We assume that the standard model gauge fields and fermions propagate in the bulk. Minimal supersymmetry in five dimensions requires eight supercharges, hence a 5D gauge multiplet consists of a 4D $\mathcal{N}=1$ vector multiplet $V$
and a chiral multiplet $\Sigma$ in the adjoint representation. To obtain a massless gauge boson, we take $V$ to be even under the $\mathbb{Z}_2$ parity $y \rightarrow -y$ and $\Sigma$ to be odd.
The massless modes, a gauge boson $A_\mu$ and a gaugino $\lambda$, have the following $y$ dependence
\cite{Gherghetta:2000qt},
\begin{equation}
\begin{split}
\\[-2.5ex]
A_\mu (x, y) = \frac{1}{\sqrt{2 \pi R}} \, A_{\mu}^{(0)} (x) + \cdots, \quad \lambda (x, y) = \frac{e^{3k |y| /2}}{\sqrt{2 \pi R}} \, \lambda^{(0)} (x) + \cdots. \\[1ex]
\end{split}
\end{equation}
Thus, the wavefunction profile of the gauge boson zero-mode is flat, whereas one can show that the corresponding KK modes are all localized toward the IR brane.
In the 4D effective theory, the gauge coupling of the zero mode is given by
\begin{equation}
\begin{split}
\\[-2.5ex]
\frac{1}{g_4^2} = \frac{2 \pi R}{g_5^2}, \\[0.5ex]
\end{split}
\end{equation}
where $g_5$ is the 5D gauge coupling (with mass dimension $-1/2$).

A bulk hypermultiplet consists of a vector-like pair of 4D $\mathcal{N}=1$ chiral multiplets, $\Psi$ and $\Psi^c$.
The bulk action is given by
\cite{Marti:2001iw}
\begin{equation}
\begin{split}
\\[-2.5ex]
S_\Psi &= \int d^5 x \biggl\{ e^{-2k |y|} \int d^4 \theta \left( \Psi^\dagger \Psi + \Psi^c {\Psi^c}^\dagger \right) \\[1.3ex]
&\quad+ e^{-3k |y|} \int d^2 \theta \,  \Psi^c \left[ \partial_y - \left( \frac{3}{2} - c_\Psi \right) k\, \epsilon (y) \right] \Psi + {\rm h.c.} \biggr\},\\[1ex]
\end{split}
\end{equation}
where we omit the gauge interactions for simplicity and
$\epsilon (y)$ is $1$ ($-1$) for positive (negative) $y$.
We assume that $\Psi$ is even under $\mathbb{Z}_2$ parity while $\Psi^c$ is odd,
which leads to the usual supersymmetric standard model matter sector below the compactification scale.
The wavefunction profile of the zero mode is controlled by the bulk mass parameter $c_\Psi$.
For $c_\Psi > 1/2$ ($c_\Psi < 1/2$), the zero mode field is localized toward the UV (IR) brane.
From the above action, the massless modes have the following $y$ dependence,
\begin{equation}
\begin{split}
\\[-2.5ex]
\Psi (x,y) = \frac{e^{- (c_\Psi - \frac{3}{2}) k |y|}}{\sqrt{ \frac{1}{\left( c_\Psi - \frac{1}{2} \right) k} \left( 1 - e^{- 2 \pi k R (c_\Psi - \frac{1}{2})} \right) }} \, \Psi^{(0)} (x) + \cdots.
\label{profile} \\[1ex]
\end{split}
\end{equation}
The 4D effective theory below the compactification scale can be obtained by substituting this expression into the action and integrating over $y$.
As in the case of the gauge fields, the wavefunctions of the KK modes are all localized toward the IR brane, regardless of the bulk mass parameter of the hypermultiplet.

\subsection{The Yukawa hierarchies}

We assume that the Higgs fields live on the IR brane. The quark and lepton fields propagate in the bulk and couple to the Higgs via brane-localized Yukawa couplings:\footnote{We introduce right-handed neutrinos for later convenience.} 
\begin{equation}
\begin{split}
\\[-2.5ex]
S_{\rm Yukawa} &= \int d^5x \, \delta(y-\pi R) \, e^{-3\pi k R} \biggl\{ \\[0.5ex]
&\qquad \int d^2 \theta \left( \tilde{y}^{ij}_u H_u Q_i \bar{u}_j + \tilde{y}^{ij}_d H_d Q_i \bar{d}_j 
+ \tilde{y}^{ij}_\nu H_u L_i \bar{\nu}_j + \tilde{y}^{ij}_e H_d L_i \bar{e}_j \right) + {\rm h.c.} \biggr\}, \\[1ex]
\end{split}
\end{equation}
where $i = 1,2,3$ labels the generation and the Yukawa couplings $\tilde{y}$ have mass dimension $-1$.
We assume that these couplings are $\mathcal{O}(1)$ in units of $k^{-1}$ with anarchic flavor structure.
Using the wavefunction profile \eqref{profile}, integrating over $y$ to remove the delta function
and canonically normalizing the Higgs fields as $H_{u,d} \rightarrow e^{\pi k R} H_{u,d}$, we find the 4D effective superpotential arising from the brane-localized interactions,
\begin{equation}
\begin{split}
\\[-2.5ex]
W^{\rm 4D}_{\rm Yukawa} &= {y}^{ij}_u H_u Q_i \bar{u}_j + {y}^{ij}_d H_d Q_i \bar{d}_j 
+ {y}^{ij}_\nu H_u L_i \bar{\nu}_j + {y}^{ij}_e H_d L_i \bar{e}_j,
\label{Yukawa} \\[1ex]
\end{split}
\end{equation}
where
\begin{equation} \label{Yukawa2}
\begin{split}
\\[-2.5ex]
{y}^{ij}_u = \tilde{y}^{ij}_u k \, \zeta_{Q_i} \zeta_{\bar{u}_j}, \quad {y}^{ij}_d = \tilde{y}^{ij}_d k \, \zeta_{Q_i} \zeta_{\bar{d}_j}, \quad
{y}^{ij}_\nu = \tilde{y}^{ij}_\nu k \, \zeta_{L_i} \zeta_{\bar{\nu}_j}, \quad {y}^{ij}_e = \tilde{y}^{ij}_e k \, \zeta_{L_i} \zeta_{\bar{e}_j}. 
 \\[1ex]
\end{split}
\end{equation}
The 4D Yukawa couplings $y$ are dimensionless.
The factor $\zeta_{\Psi}$ is given by
\begin{equation} \label{eqn:zeta}
\begin{split}
\\[-2.5ex]
\zeta_{\Psi} = \sqrt{ \frac{c_\Psi - \frac{1}{2}}{e^{2 \pi k R (c_\Psi - \frac{1}{2})} - 1}} \,, 
\end{split}
\end{equation}
so that
\begin{equation}
\begin{split}
\zeta_{\Psi} &\simeq \begin{cases}
    \sqrt{  c_{\Psi} - \frac{1}{2} }  \,\, e^{- (c_{\Psi} - \frac{1}{2}) \pi k R} & \left( c_{\Psi} \gg 1/2 \right) \\
    \\
    \frac{1}{\sqrt{ 2 \pi k R}} & \left( c_{\Psi} \sim 1/2 \right) \\
    \\
    \sqrt{ \frac{1}{2} - c_{\Psi} }  & \left( c_{\Psi} \ll 1/2 \right)
\end{cases}\label{zetaexpression} \\[1ex] 
\end{split}
\end{equation}
with an exponential suppression for $c_{\Psi} \gg 1/2$.
To explain the Yukawa hierarchies of quarks, the 1st and 2nd generations of quark multiplets are localized toward the UV brane
and they have bulk mass parameters $c_{\Psi_{1,2}} > 1/2$.
The right-handed bottom quark multiplet also lives near the UV brane and has $c_{\, \bar{d}_3} > 1/2$.
On the other hand, the 3rd generation left-handed quark multiplet and the right-handed top quark multiplet are localized toward the IR brane
and have $c_{Q_3}, c_{\, \bar{u}_3} < 1/2$.
For leptons, all generations are localized toward the UV brane with bulk mass parameters larger than $1/2$.

In the following discussion, we concentrate on the Yukawa couplings of quarks.
The similar discussion can be applied to the lepton case.
We define diagonalization matrices of the quark Yukawa matrices \eqref{Yukawa} as
\begin{equation}
\begin{split}
\\[-2.5ex]
&u = V_u \, u_0, \qquad \bar{u} = V_{\bar{u}} \, \bar{u}_0, \\[0.5ex]
&d = V_d \, d_0, \qquad \bar{d} = V_{\bar{d}} \, \bar{d}_0, \\[1ex]
\end{split}
\end{equation}
where $(u_0, \bar{u}_0, d_0, \bar{d}_0)$ are mass eigenstates of quarks
and the $V$'s are unitary matrices.
The Cabibbo-Kobayashi-Maskawa (CKM) matrix is then given by $V_{\rm CKM} = V_u^\dagger V_d$.
From the forms of the Yukawa couplings \eqref{Yukawa} and \eqref{Yukawa2},
the quark masses are approximately given by
\begin{equation}
\begin{split}
m_{u_i} \simeq \zeta_{Q_i} \zeta_{\bar{u}_i} v \sin \beta, \qquad m_{d_i} \simeq \zeta_{Q_i} \zeta_{\bar{d}_i} v \cos \beta, \label{quarkmass} \\[1ex]
\end{split}
\end{equation}
where we define $\tan \beta = \langle H_u \rangle / \langle H_d \rangle$ and the Higgs expectation value $v \simeq 174 \, \rm GeV$.
The elements of the diagonalization matrices of the Yukawa matrices, $V_u$, $V_d$, $V_{\bar{u}}$ and $V_{\bar{d}}$,
are approximately given by
\cite{Agashe:2004cp}
\begin{equation}
\begin{split}
&| \left( V_u \right)_{ij} | \simeq | \left( V_d \right)_{ij} | \simeq | \left( V_{\rm CKM} \right)_{ij} |
\simeq \frac{\zeta_{Q_j}}{\zeta_{Q_i}} \qquad \text{for \, $j \leq i$,} \\[0.5ex]
&| \left( V_{\bar{u}} \right)_{ij} | \simeq \frac{\zeta_{\bar{u}_j}}{\zeta_{\bar{u}_i}}, \qquad 
| \left( V_{\bar{d}} \right)_{ij} | \simeq \frac{\zeta_{\bar{d}_j}}{\zeta_{\bar{d}_i}} \qquad \text{for \, $j \leq i$,}
\label{diagmatrix} \\[1ex]
\end{split}
\end{equation}
where the $i$ and $j$ indices are interchanged for $j > i$.
The CKM elements can be fitted by
\begin{equation}
\begin{split}
\\[-2.5ex]
&| \left( V_{\rm CKM} \right)_{21} | \simeq \lambda, \qquad 
| \left( V_{\rm CKM} \right)_{32} | \simeq \lambda^2, \qquad | \left( V_{\rm CKM} \right)_{31} | \simeq \lambda^3, \\[1.5ex]
\end{split}
\end{equation}
where $\lambda \sim 0.2$ with a reasonable accuracy.
From the equations \eqref{quarkmass} and \eqref{diagmatrix},
we now have only two free parameters such as $\zeta_{Q_3}$ and $\tan \beta$.
Explicitly, the wavefunction factors of quarks are written in terms of these two parameters as
\begin{equation}
\begin{split}
\\[-2.5ex]
&\zeta_{Q_1}  \simeq \lambda^3 \zeta_{Q_3}, \qquad \qquad \,\,\,\,\,  \zeta_{Q_2}  \simeq \lambda^2 \zeta_{Q_3}, \\[1ex]
&\zeta_{\bar{u}_1} \simeq \frac{m_u}{\lambda^3 \zeta_{Q_3} v \sin \beta}, \qquad \zeta_{\bar{u}_2} \simeq \frac{m_c}{\lambda^2 \zeta_{Q_3} v \sin \beta},
\qquad \zeta_{\bar{u}_3} \simeq \frac{m_t}{\zeta_{Q_3} v \sin \beta}, \\[1ex]
&\zeta_{\bar{d}_1} \simeq \frac{m_d}{\lambda^3 \zeta_{Q_3} v \cos \beta}, \qquad \zeta_{\bar{d}_2} \simeq \frac{m_s}{\lambda^2 \zeta_{Q_3} v \cos \beta},
\qquad \zeta_{\bar{d}_3} \simeq \frac{m_b}{\zeta_{Q_3} v \cos \beta}. \label{wffactor} \\[1ex]
\end{split}
\end{equation}

To estimate numerical values, we use the renormalized quark masses at the $\sim 10 - 30$ TeV scale~\cite{Xing:2007fb}\footnote{These values depend on the spectrum of superpartners and the specific choice of renormalization scale, but not enough to affect our subsequent analysis. The renormalized top mass given here is valid for $\tan \beta \gsim 3$. It increases substantially at very low $\tan \beta$ due to the RG effect of the larger top Yukawa coupling.}
\begin{equation}
\begin{split} \label{eqn:TeVquarkmass}
\\[-2.5ex]
&m_u \sim 1 \, {\rm MeV}, \qquad 
m_c \sim 500 \, {\rm MeV}, \qquad m_t \sim 150 \, {\rm GeV}, \\[1ex]
&m_d \sim 2 \, {\rm MeV}, \qquad 
m_s \sim 40 \, {\rm MeV}, \qquad m_b \sim 2 \, {\rm GeV}. \\[1ex]
\end{split}
\end{equation}
For low $\tan \beta$ and $\zeta_{Q_3} \sim 1$, we have $\zeta_{\bar{u}_3} \sim 1$ and $\zeta_\Psi \lsim  \frac{1}{\sqrt{2 \pi k R}}$ for all other quark multiplets.
 Thus, by (\ref{zetaexpression}), $Q_3$ and $\bar{u}_3$ are localized toward the IR brane with $c_{Q_3, \, \bar{u}_3}<1/2$, whereas 
  the other quark multiplets are localized toward the UV brane, with $c_\Psi>1/2$. 
In section~\ref{sec:RPV}, we use these expressions to estimate the size of RPV couplings in the 4D effective superpotential.

\section{SUSY breaking} \label{sec:SUSYbreaking}

In this section, we consider SUSY breaking in the supersymmetric RS model.
We assume that a SUSY breaking sector is localized on the UV brane.
This is a natural geometric way to make stops light~\cite{Larsen:2012rq}
(and hence avoid tuning of the Higgs potential), since stops are localized near the IR brane,
away from the source of SUSY breaking. By contrast, UV-brane localized squarks and sleptons
can get soft masses via a direct (higher-dimensional) coupling to the SUSY breaking sector.
Gauginos can also couple directly to the SUSY breaking sector, but the resulting gaugino mass
is often strongly suppressed relative to the scalar mass, as we explain below.
In this case, the UV-brane localized scalars must be very heavy to reproduce reasonable, weak-scale gaugino masses, leading to a ``natural SUSY'' spectrum~\cite{naturalSUSYspectrum, Papucci:2011wy}. 

We refer to this class of models, with an IR-brane-localized Higgs, hierarchical Yukawa couplings generated by the wavefunction profiles of bulk fermions as in~\S\ref{sec:SUSYRS}, and UV-brane-localized SUSY breaking leading to decoupled first and second generation squarks, as ``warped natural SUSY.'' Models of this type have been considered previously in~\cite{Partly,Sundrum:2009gv,Redi:2010yv,Gherghetta:2011wc}.\footnote{In~\cite{Partly} the gauginos are decoupled as well as the first and second generation squarks.}

\subsection{Couplings to the SUSY breaking sector}

Gaugino, squark, and slepton masses can be generated by higher-dimensional operators on the UV brane. The leading contributions are:
\begin{equation}
\begin{split}
\\[-2.5ex]
S_{\rm UV} \supset \int d^5 x \, \delta(y)  \left[  c^i_{\; j} \int d^4 \theta \,  \frac{X^\dagger X}{k M^2}
\Psi^\dagger_i \Psi^j+b^{a b} \int d^2 \theta \, \frac{X}{k M}  \, \tr \, W^\alpha_a W_{\alpha\; b} + {\rm h.c.} \right]\,, \\[1ex]
\end{split}
\end{equation}
where $X \equiv \theta^2 F$ is a SUSY breaking spurion with nonzero $F$-term, $M$ is the mediation scale,
and in general the coefficients $c^i_{\;j}$ and $b^{a b}$ are constrained only by gauge invariance,
with $\mathcal{O}(1)$ values and an otherwise anarchic structure. This gives the gaugino and scalar masses:
\begin{equation}
\begin{split}
m_{\lambda}^{a b} = \frac{F}{\pi k R M} b^{a b} \,, \qquad (m^2)^i_{\; j} = \frac{F^2}{M^2} \eta_i \eta_j c^i_{\; j} \,, \\[1ex] \label{eqn:squarkmass}
\end{split}
\end{equation}
where
\begin{equation} \label{eqn:etaexact}
\eta_{\Psi} = \sqrt{ \frac{c_\Psi-\frac{1}{2}}{1-e^{-2 \pi k R (c_\Psi-\frac{1}{2})}}} \;,
\end{equation}
so that
\begin{equation}
\begin{split}
\\[-2.5ex]
\eta_{\Psi} &\simeq \begin{cases}
    \sqrt{  c_{\Psi} - \frac{1}{2} } \simeq \sqrt{ \frac{1}{\pi k R} \log \zeta_\Psi^{-1} } & \left( c_{\Psi} \gg 1/2 \right) \\
    \\
    \frac{1}{\sqrt{ 2 \pi k R}} \simeq \zeta_\Psi & \left( c_{\Psi} \sim 1/2 \right) \\
    \\
    \sqrt{ \frac{1}{2} - c_{\Psi} } \,\, e^{- (\frac{1}{2} - c_{\Psi}) \pi k R} \simeq \zeta_\Psi e^{- \pi k R \, \zeta_\Psi^2} & \left( c_{\Psi} \ll 1/2 \right)
\end{cases}\label{etaexpression1} \\[2ex] 
\end{split}
\end{equation}
with $\zeta_\Psi$ given by~(\ref{zetaexpression}). Thus, the soft masses for $Q_3, \bar{u}_3$ are exponentially suppressed,
and other sources of SUSY breaking such as loop corrections induced by the other soft masses will dominate.
For the other squarks (at low $\tan \beta$ and $\zeta_{Q_3} \sim 1$) we have $2 \lsim \log \zeta^{-1}_{\Psi} \lsim 7$, hence
\begin{equation}
\begin{split}
m^2_{\tilde{q}} \, \gsim \, \frac{2 F^2}{\pi k R M^2} \,. \\[1.5ex]
\end{split}
\end{equation}
Comparing with the gaugino masses, we conclude that the UV brane localized scalars are an order of magnitude or more heavier than the gauginos in general.
Compatibility with the experimental lower bound on the gluino mass suggests that these scalars are above $10$ TeV,
hence they are absent from the low energy effective theory.\footnote{By adjusting $\zeta_{Q_3}$ and $\tan \beta$,
we can make some scalars lighter at the expense of making others heavier.}

The mass hierarchy between the UV-brane-localized scalars and the gauginos can be much larger when
the SUSY breaking sector contains no singlets with large $F$-terms, i.e.\
when the SUSY breaking spurion $X$ is charged under some symmetry of the SUSY breaking sector.
In this case, the leading contribution to gaugino masses is:
\begin{equation}
\begin{split}
\\[-2.5ex]
S_{\rm UV} \supset \int d^5 x \,\delta(y) \left[b^{a b} \int d^4 \theta \, \frac{X^\dagger X}{k M^3}  \, \tr \, W^\alpha_a W_{\alpha\; b} + {\rm h.c.}\right]\,, 
\end{split}
\end{equation}
giving
\begin{equation} \label{eqn:gauginoMassDSB}
\begin{split}
m_{\lambda}^{a b} = \frac{F^2}{\pi k R M^3} b^{a b}\,. \\[1.5ex]
\end{split}
\end{equation}
For a high messenger scale, $F/M^2 \ll 1$, and therefore $m_\lambda \ll m_{\tilde{q}}$, and the UV-brane-localized scalars decouple.\footnote{The corresponding KK modes do not decouple, however.}

At first sight, this situation appears to be dangerous for naturalness. In the MSSM, decoupling the scalars leads to quadratic divergences in the Higgs soft masses.
However, in our model the Higgs fields are localized on the IR brane, and hence the one-loop radiative corrections
to the Higgs soft masses are cut off at the IR scale $\Lambda_{\rm IR} \sim \pi k'$:\footnote{
See~\S\ref{subsec:stopHiggsMass} for a discussion of two-loop quadratic divergences induced by the gauge interactions.}
\begin{equation}
\begin{split}
\\[-2.5ex]
\Delta m_H^2 \simeq - \frac{3}{8 \pi^2} \, y_\Psi^2 \, \Lambda_{\rm IR}^2 \,. \\[1ex]
\end{split}
\end{equation}
Because of the small Yukawa couplings for the other quarks and leptons, only the stop needs to be
in the effective theory~\cite{Sundrum:2009gv,Gherghetta:2011wc},
and the mass scale of the UV brane localized squarks and sleptons has little effect on tuning (see, however, \S\ref{sec:unification}).

In fact, if the UV-brane couplings $c^i_{\;j}$ are flavor anarchic, then to avoid excessive flavor
and CP violation we require $m_{\tilde{q}} \gsim 5 \times 10^4$ TeV,\footnote{Flavor violation
is communicated to the light quarks via $\alpha_s$ suppressed squark-gluino loops,
leading to a weaker bound than that for generic new physics~\cite{bonatalk}.} or
\begin{equation}
\begin{split}
\\[-2.5ex]
\frac{F}{M} \gsim 2\times 10^5\; \mathrm{TeV} \\[1ex]
\end{split}
\end{equation}
which is compatible with TeV scale gauginos if $F/M^2 \lsim 2\times 10^{-4}$.

\subsection{Anomaly and radion mediation}

Since gaugino masses can be strongly suppressed relative to the squark masses,
it is important to consider other ways in which SUSY breaking can be mediated to the fields
in the low energy effective theory. In this subsection, we consider the effects of anomaly and radion mediation.
An $F$-term for the radion superfield $T$ leads to gaugino masses at tree level~\cite{Chacko:2000fn} (see also \cite{Kobayashi:2000ak}).
The kinetic term for the gauge zero mode in the 4D effective Lagrangian is given by
\begin{equation}
\begin{split}
\Delta \mathcal{L}_4 = \frac{1}{2g_5^2} \int d^2 \theta \, T \, \tr \, W^\alpha W_\alpha + {\rm h.c.}, \\[1ex]
\end{split}
\end{equation}
where $T$ is normalized so that $\langle T \rangle = \pi R + \theta^2 F_T$.
The radion $F$-term generates a gaugino mass,
\begin{equation}
\begin{split}  \label{eqn:RadMedTree}
m_{\lambda} = \frac{F_T}{2T}\,. \\[1ex]
\end{split}
\end{equation}
Notably, if tree-level radion mediation dominates, then the gaugino masses are degenerate at leading order.

However, radion mediation also occurs at one-loop level, along with anomaly mediation~\cite{anomaly}.
The calculation of these contributions is somewhat complicated due to the warped extra dimension and the hard SUSY breaking implied by the decoupling of the heavy scalars.

We make a rough estimate as follows. We first consider the case of pure anomaly mediation, where $F_T=0$. 
We work in the dual four-dimensional description, where the standard model is a weakly-gauged flavor symmetry of a strongly interacting SCFT.
Before decoupling the fundamental scalars (dual to the UV brane localized scalars), SUSY is softly broken and the usual anomaly-mediation formulas for the gaugino masses apply. The gaugino masses will differ from their values in the minimal supersymmetric standard model (MSSM) due to contributions to the beta functions from CFT states, but these contributions are removed when these states are integrated out supersymmetrically due to threshold corrections to the gaugino masses, and the result is UV-insensitive.

Upon integrating out the heavy scalars while retaining their massless superpartners there is no analogous threshold correction, since the gaugino masses are protected by an R-symmetry. Thus, just below this scale the gaugino masses mimic the standard anomaly-mediated formula, counting the incomplete fermion multiplets as whole multiplets for this purpose. However, the physics is no longer UV-insensitive: the one-loop beta functions for the gaugino masses will depend only on the spectrum of complete multiplets, whereas the one-loop beta functions for the gauge couplings will incorporate all multiplets, and the anomaly mediation formula will break down under RG flow.
The resulting physics is somewhat complicated, but we expect that these RG effects will not drastically affect the gaugino spectrum. Hence, we assume that the usual anomaly mediation formula with incomplete fermion multiplets counted as whole multiplets is approximately valid at the confinement scale. Improving and/or validating this approximation is left to a future work.



Next, we consider the case of pure radion mediation, where $F_\phi=0$ for the superconformal compensator $\phi$~\cite{compensator}. The warp factor superfield $\omega \equiv \phi e^{-k T}$ is the effective superconformal compensator on the IR brane
\cite{Luty:2002ff}, hence there is a one-loop radion-mediated contribution to the gaugino masses given by replacing $F_\phi$ with $F_\omega / \omega$ in the usual anomaly mediated formula. This replacement affects only the IR-brane localized fields, whose scalar partners are not decoupled, hence there is no subtlety with hard SUSY breaking in this case.


Combining these two results, we estimate
\begin{equation} \label{eqn:ARMedLoop}
\begin{split}
\\[-2.5ex]
&m_{\tilde{g}}^{(1)} \, \sim \, - \frac{9}{2}
\frac{g_3^2}{16 \pi^2} F_\phi + \frac{3}{2} \frac{g_3^2}{16 \pi^2} \frac{F_{\omega}}{\omega}, \\[2ex]
&m_{\widetilde{W}}^{(1)} \, \sim \, - \frac{3}{2}
\frac{g_2^2}{16 \pi^2} F_\phi + \frac{5}{2} \frac{g_2^2}{16 \pi^2} \frac{F_{\omega}}{\omega}, \\[2ex]
&m_{\widetilde{B}}^{(1)} \, \sim \, \frac{51}{10} \frac{g_1^2}{16 \pi^2} F_\phi
+ \frac{3}{2} \frac{g_1^2}{16 \pi^2} \frac{F_{\omega}}{\omega}, \\[2ex]
\end{split}
\end{equation}
where the different running of the gaugino masses and gauge couplings below the heavy scalar threshold is neglected as explained above. This and other effects from radion stabilization may lead to a substantial model dependence in the gaugino masses, but we use the heuristic estimates~(\ref{eqn:ARMedLoop}) for the remainder of the paper for want of a better calculation, which we leave to a future work. Fine tuning will depend primarily on a quadratic divergence which we discuss later, so our conclusions will not depend on the detailed coefficients in~(\ref{eqn:ARMedLoop}).

Since
\begin{equation}
\begin{split}
\\[-2.5ex]
\frac{F_{\omega}}{\omega} = F_{\phi} -(\pi k R) \frac{F_T}{T} \\[1.5ex]
\end{split}
\end{equation}
receives a volume-factor ($\pi k R$) enhancement, the one-loop radion-mediated contribution to the gluino mass
is similar in size to the tree-level contribution~(\ref{eqn:RadMedTree}).
The gaugino masses are the sum of these two contributions.
Typically, in models where SUSY is broken dynamically in a hidden sector,
$|F_\phi| \simeq m_{3/2}$,\footnote{See~\cite{splitSUSY} for some examples where this relationship fails.}
implying an upper bound on the gravitino mass $m_{3/2} \lsim 100 \, \rm TeV$ for TeV scale gauginos.


Anomaly and radion mediation also contribute to the scalar masses. There is a tree-level radion mediated contribution~\cite{Marti:2001iw}:
\begin{equation}
\begin{split}
\\[-2.5ex]
m_{\Psi} = \left| \frac{(1/2 - c_{\Psi}) k \pi R}{2 \sinh \left[(1/2 - c_{\Psi}) k \pi R \right]} \right| \frac{F_T}{T} \,. \\[1.5ex]
\end{split}
\end{equation}
However, this is exponentially suppressed unless $c_\Psi \simeq 1/2$, and is therefore negligible in most cases.
Anomaly mediation and radion mediation do contribute to scalar masses at one-loop.
Nonetheless, this is only relevant for the stop mass and the Higgs mass (discussed in \S\ref{subsec:stopHiggsMass}), since the UV brane localized scalars
acquire a much larger mass, as argued above.

\subsection{A coincidence problem}

A generic problem with models which extend the MSSM with additional new physics at the TeV scale is that this typically requires a nontrivial coincidence between the supersymmetry breaking scale and the scale associated to the MSSM extension~\cite{Randall:2012dm}, such as a mass scale or a confinement scale associated to the new physics. In the present context, this coincidence manifests itself as the confluence between the gaugino masses and the compactification scale. If the gaugino masses are set by~(\ref{eqn:gauginoMassDSB}), then this confluence seems to be hard to explain, as the suppression factor $F/M^2$ bears no obvious relation to the compactification scale.

On the other hand, if the gaugino masses are generated by radion mediation (potentially with an anomaly-mediated contribution as well), then the gaugino masses are set by the same physics which stabilizes the radion
(see e.g.~\cite{radion,Luty:2000ec}) and sets the compactification scale. In this case, it is conceivable that a well-designed mechanism of radion stabilization could explain the coincidence between these two scales. Motivated by this, we assume henceforward that the radion and anomaly mediated contributions, (\ref{eqn:RadMedTree}, \ref{eqn:ARMedLoop}), dominate over direct couplings to the SUSY breaking sector. 
However, we know of no concrete model which fully explains this coincidence, and we leave the problem for a future work.

The coincidence problem further motivates our choice of heavy first and second generation squarks. If by some mechanism the large squark masses~(\ref{eqn:squarkmass}) were forbidden, we could build a model with first and second generation squarks in the low-energy effective theory. However, in this case the principle contributions to the squark masses would come from gaugino mediation and/or anomaly and radion mediation, as detailed in the next subsection. These contributions are the same for the stop and other squarks, hence their masses would be similar, requiring light first and second generation squarks in conflict with LHC searches.\footnote{See however~\cite{Graham:2014vya}.} The only way to introduce a substantial splitting
is by coupling the first two generations directly to the SUSY breaking sector, as above. However, this requires a further coincidence to ensure a TeV-scale splitting; a larger splitting would remove the squarks from the effective theory, as before, whereas a smaller splitting would have little effect.


\subsection{Light scalars and fine tuning} \label{subsec:stopHiggsMass}

Since the up-type Higgs soft mass $m_{H_u}^2$ is radiatively generated by the stop masses,
in general light stops are needed for natural electroweak symmetry breaking.
At present, LHC searches place a strong constraint on the stop mass in R-parity conserving models~\cite{stopbound}. In a significant fraction of parameter space, the bound is $m_{\tilde{t}} \gsim 650 \, \rm GeV$.
While baryonic RPV can significantly reduce this limit,
Higgs coupling measurements still constrain the light stop masses~\cite{Fan:2014txa}.
In a typical NMSSM-like model, for the $h \rightarrow g g$ and $h \rightarrow \gamma \gamma$ couplings to agree with their presently measured values,
one stop mass has to be at least $300 \, \rm GeV$,
which is still consistent with naturalness.

There are three important contributions to the stop mass, two of which are positive and one of which is negative.
Firstly, anomaly and radion mediation generate non-zero stop masses at the compactification scale~\cite{Luty:2002ff},
\begin{equation} \label{eqn:anomRadStop}
\begin{split}
\\[-2.5ex]
m_{\tilde{t}(0)}^2 = \frac{1}{(16\pi^2)^2} \left| \frac{F_\omega}{\omega} \right|^2 \left( 8 g_3^4 + \cdots \right).  \\[1.5ex]
\end{split}
\end{equation}
where the warp factor superfield $\omega \equiv \phi e^{-kT}$ replaces the superconformal compensator $\phi$ because the stops
are localized toward the IR brane
and their effective cutoff scale is set by $\omega$.

Secondly, positive squark masses
are radiatively generated by the gluino below the cutoff $\Lambda_{\rm IR} \sim \pi k'$,
\begin{equation} \label{eqn:gluinoStop}
\begin{split}
\delta m_{\tilde{t}}^2 \, |_{\rm gluino} \simeq \, \frac{8\alpha_3}{3 \pi} |m_{\tilde{g}}|^2 \left[ \log \frac{\Lambda_{\rm IR}}{m_{\tilde{g}}} +\frac{1}{2} \right]\,, \\[1ex]
\end{split}
\end{equation}
where the second term is a threshold correction~\cite{Pierce:1996zz} evaluated in the limit $m_{\tilde{t}} \ll m_{\tilde{g}}$, which we include for completeness due to the small log inherent in a low cutoff. The scenario where~(\ref{eqn:gluinoStop}) is the dominant contribution to the stop mass is known as ``gaugino mediation''~\cite{gaugino}.

The gluino mass is stringently constrained by the LHC experiments.
Without RPV, gluino pair production leads to copious top quarks, high-$p_T$ ($b$-)jets and missing transverse energy (MET) from stable lightest supersymmetric particles (LSPs).
This gives a stringent lower bound on the gluino mass, approaching the kinematic limit at Run I of the LHC ($\sim 20 \, {\rm fb}^{-1}$ at $8 \, \rm TeV$)
which corresponds to $m_{\tilde{g}} \sim 1.2 - 1.4 \, \rm TeV$.
With RPV, there is a sharp reduction of missing energy, which can relax the limit from MET-based searches.
However, in this case, the extra jets replace missing energy, which gives rise to the limit on the gluino mass from searches that do not require MET.
The lower bound on the gluino mass is then at least $1 \, \rm TeV$
\cite{Evans:2013jna,ATLAS-CONF-2013-091}.

In either case, the gluino mediated contribution to the stop mass is sizable.
However, if the gluino mass is generated by anomaly/radion
mediation, then~(\ref{eqn:anomRadStop}) is typically dominant over the gaugino-mediated contribution, unless $|F_\omega/\omega| \ll |F_\phi|$, which requires $F_T/T \sim \frac{1}{\pi k R} F_\phi$. This occurs naturally in the Luty-Sundrum mechanism of radion stabilization~\cite{Luty:2000ec}, where the gaugino-mediated contribution dominates.

\begin{figure}[!t]
  \begin{center}
  \vspace{-1cm}
          \includegraphics[clip, width=10cm]{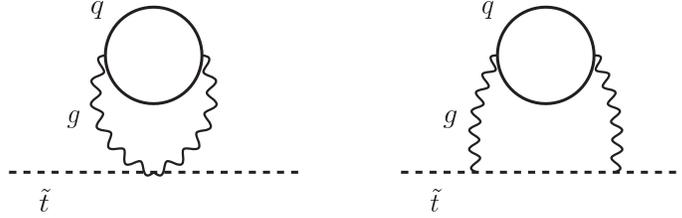}
    \caption{Two-loop diagrams contributing to a quadratic divergence in the stop mass induced by the absence of the UV-brane localized squarks from the effective theory.}
    \label{fig:divergence}
  \end{center}
\end{figure}

In addition to these positive contributions, there is an important negative contribution to the stop masses as well. This comes from a two-loop quadratic divergence induced by the light quarks, shown in figure~\ref{fig:divergence}. Such a divergence is possible because the low-energy effective theory contains incomplete quark multiplets with missing squarks, due to the decoupling of the UV-brane localized squarks.

While quadratic divergences do not occur in a mass-independent scheme such as $\overline{\rm DR}$, large threshold corrections can appear in these schemes, and they play a similar role. For instance, if the UV theory is supersymmetric, then the threshold correction upon integrating out a heavy scalar can be interpreted as roughly equivalent to the quadratic divergence which would appear in the low energy effective theory in another scheme.

Consider a gauge theory with two charged chiral multiplets in representations $r$ and $R$. Suppose that the fermionic components of these multiplets are massless and that scalar masses are $m$ and $M$, respectively, where $m \ll M$. The threshold correction to $m$ upon integrating out the heavy scalar at a scale $\mu$ takes the form~\cite{Poppitz:1996xw,Hisano:2000wy}:
\begin{equation}
\begin{split}
\\[-2.5ex]
\Delta m^2(\mu) = - 4 \left(\frac{g^2}{16 \pi^2}\right)^2 C(r) S(R) M^2 \left(\frac{\pi^2}{3}-2-\log \frac{M^2}{\mu^2}\right) \label{eqn:quadDiv} \\[1ex]
\end{split}
\end{equation}
in the $\overline{\rm DR}'$ scheme,\footnote{There is some disagreement about this result in the literature. For instance~\cite{Agashe:1998zz} finds an extra term $\log 4 \pi - \gamma$ within the parentheses. While~\cite{Poppitz:1996xw,Hisano:2000wy,Agashe:1998zz} all claim to use $\overline{\rm DR}'$, this may be due to some subtle residual scheme dependence.} where $C(r)$ denotes the quadratic Casimir of $r$ (normalized to $\frac{N^2-1}{2 N}$ for $SU(N)$) and $S(R)$ denotes the Dynkin index of $R$ (normalized to $\frac{1}{2}$ for $SU(N)$). The coefficient of the log is fixed to agree with the two-loop beta functions in this scheme~\cite{Martin:1993zk}.

Taking $\Delta m^2(M)$ as an estimate of the quadratic divergence, we find for the stop:
\begin{equation}
\begin{split}
\\[-2.5ex]
\delta m_{\tilde{t}}^2 \, |_{\rm 2-loop} \simeq - \frac{3}{2\pi^2} \left(\frac{\pi^2}{3}-2\right) \alpha_3^2 \, \Lambda_{\rm IR}^{2} \, . \\[1ex]
\end{split}
\end{equation}
Because of this negative contribution, there is an upper limit on the cutoff $\Lambda_{\rm IR}$ relative to the gluino mass $m_{\tilde{g}}$ to ensure a non-tachyonic stop. For instance, in the case $F_{\omega}/\omega \ll F_\phi$, as in the Luty-Sundrum model, we find $m_{\tilde{g}} \gsim \Lambda_{\rm IR}/12$. This limit is relaxed when $F_\omega/\omega$ is substantial. In either case, there is some degree of cancellation between the quadratic divergence and anomaly and radion-mediated effects. So long as the stop is not tachyonic, this cancellation is only important insofar as it leads to tuning in the Higgs potential, which we quantify below.

We now discuss contributions to the up-type Higgs mass. As with the stop, there is a direct contribution from anomaly/radion mediation:
\begin{equation} \label{eqn:HiggsM0}
\begin{split}
\\[-2.5ex]
m_{H_u(0)}^2 = \frac{1}{(16\pi^2)^2} \left| \frac{F_\omega}{\omega} \right|^2 \left( -16 y_t^2 g_3^2 + 18 y_t^4 + \cdots \right). \\[1ex]
\end{split}
\end{equation}
The up-type Higgs mass also receives one-loop corrections from the stop and wino masses, as well as a quadratically divergent correction of the form~(\ref{eqn:quadDiv}). We separate these corrections into two pieces. Firstly, we obtain corrections proportional to $\Lambda_{\rm IR}^2$ from the quadratic divergences in the stop and up-type Higgs masses:
\begin{equation}
\begin{split}
\\[-2ex]
\delta m_{H_u}^2 \, |_{\rm quad} \, \simeq \, \left[ \, \frac{9 y_t^2 \alpha_3^2}{8 \pi^4} \log \frac{\Lambda_{\rm IR}}{m_{\tilde{t}}}
- \frac{27 \alpha_2^2}{32 \pi^2} \, \right] \left(\frac{\pi^2}{3}-2\right) \Lambda_{\rm IR}^2 \, , \label{eqn:HiggsM1} \\[1.5ex]
\end{split}
\end{equation}
where we neglect $g_1,g_2 \ll g_3$ in the first term and $g_1 \ll g_2$ in the second term. The remaining contributions to $m_{H_u}^2$ are all proportional to $|F_\phi|^2$ or $|F_\omega/\omega|^2$:
\begin{equation}
\begin{split}
\\[-2.5ex]
\delta m_{H_u}^2 \, |_{\rm A/R} \, &\simeq \, m_{H_u(0)}^2 - \frac{3 y_t^2}{4 \pi^2} m_{\tilde{t}(0)}^{2} \log \frac{\Lambda_{\rm IR}}{m_{\tilde{t}}} + \frac{3 \alpha_2}{2 \pi} |m_{\tilde{W}}|^2 \log \frac{\Lambda_{\rm IR}}{m_{\tilde{W}}} \\[1ex]
&\qquad- \frac{2 \alpha_3 y_t^2}{\pi^3} |m_{\tilde{g}}|^2 \left(\frac{1}{2} \log^2 \frac{\Lambda_{\rm IR}}{m_{\tilde{g}}}
+\log \frac{\Lambda_{\rm IR}}{m_{\tilde{g}}} \log \frac{m_{\tilde{g}}}{m_{\tilde{t}}} \right) \,, \label{eqn:HiggsM2} \\[1.5ex]
\end{split}
\end{equation}
where we include~(\ref{eqn:HiggsM0}), $m_{\tilde{t}(0)}^{2}$ is given by~(\ref{eqn:anomRadStop}), and we neglect threshold corrections. The $\alpha_3$-dependent corrections in (\ref{eqn:HiggsM1}) and (\ref{eqn:HiggsM2}) arise from radiative corrections to the stop mass which are communicated to the Higgs mass through the one-loop renormalization group equations.

Unless the cutoff $\Lambda_{\rm IR}$ is naturally related to $|F_\phi|$ and/or $|F_\omega/\omega|$ in a particular way, the quadratic divergence~(\ref{eqn:HiggsM1}) will be an irreducible source of fine tuning. For a sufficiently high cutoff, the dominant constraint on $m_{\tilde{g}}$ is that of a non-tachyonic stop, rather than the direct LHC constraint. In this case, (\ref{eqn:HiggsM1}) is typically the dominant source of tuning.
We estimate the fine tuning in this case using the fine-tuning measure~\cite{Kitano:2006gv}:
\begin{equation} 
\begin{split}
\\[-2.5ex]
\Delta \equiv \frac{2\, \delta m_{H_u}^2}{m_h^2}, \label{eqn:finetuning} \\[1ex]
\end{split}
\end{equation}
for $\tan \beta \gsim 2$, where $m_h = 125 \, \rm GeV$ is the observed Higgs boson mass. The result depends strongly on the cutoff. For example, with $\Lambda_{\rm IR} = 15$ TeV we find $\Delta \sim 5$, corresponding to $20\%$ fine tuning. To obtain this result, we substitute the running gauge couplings evaluated at the cutoff into (\ref{eqn:HiggsM1}), where we assume the spectrum $m_{\tilde{g}} = m_{\tilde{W}} = 1.5$ TeV, $m_{\tilde{t}} = 500$ GeV, and $m_{\tilde{h}} = 200$ GeV, giving $\alpha_3(15\ \mathrm{TeV}) \simeq 0.076$ and $\alpha_2(15\ \mathrm{TeV}) \simeq 0.032$. The degree of fine tuning depends only weakly on the superpartner spectrum, but increases rapidly as the cutoff is raised; for $\Lambda_{\rm IR} = 30$ TeV, we obtain $\Delta \simeq 25$, corresponding to $4\%$ tuning.


When the cutoff is low, the quadratic divergences are subdominant, and the degree of tuning is more difficult to characterize in general. The LHC constraints on the gluino mass are likely to be the dominant source of tuning, but estimating this tuning requires a precise estimation of the anomaly and radion mediated contributions to the gaugino, stop, and Higgs masses.
 As one example, we consider the case $|F_\omega/\omega| \ll |F_\phi|$, so that $m_{H_u(0)}^2$ and $m_{\tilde{t}(0)}^2$ are negligible. Assuming the validity of~(\ref{eqn:RadMedTree}) and (\ref{eqn:ARMedLoop}), we find that $|m_{\tilde{g}}| \sim |m_{\tilde{W}}|$.\footnote{This is a numerical accident, since the wino mass comes mainly from tree-level radion mediation, (\ref{eqn:RadMedTree}), whereas the gluino mass comes mainly from the one-loop contribution, (\ref{eqn:ARMedLoop}). Model-dependence in (\ref{eqn:ARMedLoop}) may alter this conclusion.} The gaugino-mediated terms in~(\ref{eqn:HiggsM2}) scale with $|F_\phi|^2$, and provide a measure of fine tuning. For instance, for $m_{\tilde{g}} = m_{\tilde{W}} = 1.5$ TeV and $\Lambda_{\rm IR} = 10$ TeV, we obtain $m_{\tilde{t}} \simeq 500$ GeV with $\Delta \sim 2.5$, corresponding to $40\%$ (almost negligible) tuning.\footnote{The degree of tuning is decreased by a fortuitous partial cancellation between the gluino- and wino-mediated terms.} In this regime, fine tuning increases rapidly as the gluino mass is raised, but depends only weakly on the cutoff.
 

Thus, a natural model is possible if we can achieve a low cutoff with light gauginos. Improved limits on the gluino mass and/or direct or indirect constraints on the compactification scale will inevitably increase the tuning.




\section{Higgs physics} \label{sec:Higgs}

We comment briefly on the Higgs sector in our scenario. We employ an RS realization of the NMSSM.
To generate the $\mu$ term and explain the observed Higgs boson mass of $125 \, \rm GeV$,
we introduce a bulk singlet field $S$ localized toward the IR brane
together with a superpotential on the IR brane:
\begin{equation}
\begin{split}
\\[-2.5ex]
S_{\rm Higgs} = \int d^5x \, \delta(y-\pi R) \, e^{-3k\pi R} \left[ \int d^2 \theta \left( \tilde{\lambda}
S H_u H_d + \frac{\tilde{\kappa}}{3} S^3 \right) + {\rm h.c.}  \right], \\[1ex]
\end{split}
\end{equation}
where $\tilde{\lambda}$ and $\tilde{\kappa}$ are coupling constants with mass dimension $-1/2$ and $-3/2$ respectively.
We impose a $\mathbb{Z}_3$ symmetry explicitly broken on the UV brane, under which $\Phi \rightarrow e^{2\pi i /3} \, \Phi$ for every chiral multiplet in the theory. This corresponds to an accidental $\mathbb{Z}_3$ symmetry in the CFT description.

Using the wavefunction profile \eqref{profile} and integrating over $y$, we obtain the 4D effective superpotential
\begin{equation}
\begin{split}
W_{\rm Higgs}^{\rm eff} = \lambda_{\rm eff} S H_u H_d + \frac{\kappa_{\rm eff}}{3} S^3,
\label{NMSSM1} \\[1.5ex]
\end{split}
\end{equation}
where $\lambda_{\rm eff} \equiv \tilde{\lambda} \, k^{1/2} \,  \zeta_S$ and $\kappa_{\rm eff} \equiv \tilde{\kappa} \, k^{3/2} \, \zeta_S^3$ are dimensionless coupling constants, with
 $\zeta_S$ given by~\eqref{zetaexpression}.
 The coupling $\lambda_{\rm eff}$ contributes to the Higgs quartic coupling at tree-level.
The tree-level upper bound on the mass of the lightest neutral CP-even Higgs boson is then~\cite{NMSSM}
\begin{equation}
\begin{split}
m_h^2 \leq m_Z^2 \left( \cos^2 2\beta + \frac{2 \lambda_{\rm eff}}{g_1^2 + g_2^2} \sin^2 2 \beta \right).
\label{NMSSM2} \\[2ex]
\end{split}
\end{equation}
To obtain the observed Higgs mass, small $\tan \beta$ and order-one $\lambda_{\rm eff}$ are required.
Since the running of the 4D effective coupling $\lambda_{\rm eff}$ is cut-off at the IR scale,
 this coupling can be large at the electroweak scale without encountering a Landau pole~\cite{Barbieri:2006bg}.
Therefore, a $125 \, \rm GeV$ Higgs boson is easily obtained.

An explicit $\mu$ term is forbidden by the $\mathbb{Z}_3$ symmetry. Instead, to generate a Higgsino mass
 $S$ must obtain a nonzero vev, $\mu_{\rm eff} = \lambda_{\rm eff} \langle S \rangle \sim 200 \, \rm GeV$,
spontaneously breaking the $\mathbb{Z}_3$ symmetry. As a result, domain walls will appear during the electroweak phase transition in the early universe.
To avoid upsetting the balance of light elements in the universe, these domain walls must disappear before nucleosynthesis,
which requires a sufficiently large splitting between the vacuum energies of the three vacua~\cite{domainwall},
\begin{equation}
\begin{split}
\\[-2.5ex]
\Delta V \, \gsim \, \left( 1 \, {\rm MeV} \right)^4. \label{domainpotential} \\[1ex]
\end{split}
\end{equation}
This can be generated by coupling $S$ to the SUSY breaking sector on the UV brane, where the $\mathbb{Z}_3$ symmetry is explicitly broken.
For example, a tadpole can be generated:
\begin{equation}
\begin{split}
\\[-2.5ex]
V_{\not{\mathbb{Z}_3}} \, = \, a_S S +a_S^{\ast} S^{\ast}+\ldots, \\[1ex]
\end{split}
\end{equation}
where $a_S$ has mass dimension $3$.
The mass scale of this potential can be controlled by adjusting $\eta_S \simeq e^{- \left( \frac{1}{2} - c_S\right) \pi k R}$, with little effect on $\zeta_S$ (hence $\lambda_{\rm eff}$). The domain wall problem is solved without fine tuning for a wide range of $\eta_S$ values corresponding to $(20\ \mathrm{KeV})^3 \lsim a_S \lsim (200\ \mathrm{GeV})^3$.

Finally, we briefly comment on a possible solution to the strong CP problem based on an invisible axion model,
which introduces a SM singlet with a $U(1)$ Peccei-Quinn (PQ) charge and a large vev
(for a review, see e.g.~\cite{Peccei:2006as}). The phase of the singlet is the axion, whereas the vev sets the axion decay constant, and must be much larger than the electroweak scale to avoid experimental constraints.
Popular invisible axion models include the KSVZ model~\cite{KSVZ}, with additional heavy PQ-charged vector-like quarks and vanishing PQ charge for the light fields, and the DFSZ model~\cite{DFSZ}, with no additional heavy particles, two Higgs doublets, and PQ-charged standard model fields.

Since the DFSZ model only requires an additional singlet (the second Higgs doublet already being required by supersymmetry), this model is more minimal, and we focus on it here.
In this model, quarks, leptons, and the Higgs doublet transform nontrivially under the PQ symmetry, but
only the Higgs field couples to the axion singlet at the renormalizable level.
A supersymmetric extension of the model is possible~\cite{Rajagopal:1990yx}, with a superpotential
\begin{equation}
\begin{split}
\\[-2.5ex]
W_{\rm axion} \, = \, \xi A H_u H_d \, , \label{axion} \\[1ex]
\end{split}
\end{equation}
where $A$ is the axion chiral superfield
and $\xi$ is a dimensionless coupling constant.
The axion is the phase of the scalar component of $A$, which gets a large vev $\langle A \rangle \sim 10^{12} \, \rm GeV$.
Since this vev generates an effective $\mu$ term,
we require  $\xi \lsim 10^{-10}$ to preserve naturalness.
Although technically natural, there is no reason for such a small coupling in the 4D model given in~\cite{Rajagopal:1990yx}.
By contrast, the SUSY RS model provides a natural explanation if $A$ comes from
a bulk multiplet localized toward the UV brane, so that the 4D effective superpotential~\eqref{axion}
includes a wavefunction suppression factor.
In this scenario, cosmological dark matter could consist of the axion and/or its superpartner, the axino.

\section{R-parity violation} \label{sec:RPV}

\begin{figure}[t!]
\vspace{-0.5cm}
  \begin{center}
          \includegraphics[clip, width=5cm]{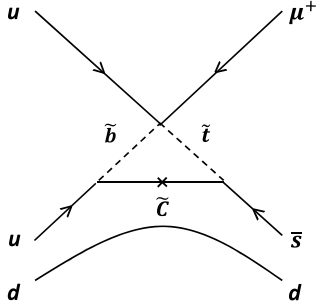}
\vspace{0cm}
    \caption{A proton decay diagram induced by the operator $\frac{1}{\Lambda} Q^3L$ and involving only sparticles which must be light to preserve naturalness. A low cutoff $\Lambda \sim 10$ TeV is ruled out.
}
    \label{fig:protondecay}
  \end{center}
\end{figure}

Randall-Sundrum models generically have a proton decay problem due to their low cutoff (see e.g.~\cite{Gherghetta:2000qt}). For instance, in the supersymmetric RS model the extremely dangerous operator $\frac{1}{\Lambda_{\rm IR}} QQQ L$ can be generated on the IR brane with $\Lambda_{\rm IR}$ near the TeV scale. The flavor-dependent wavefunction suppression implicit in the model is not enough to suppress proton decay.\footnote{This operator is dangerous even in the MSSM with the cutoff near the Planck scale, though in this case a reasonable ``Yukawa-like'' flavor structure can eliminate the issue. The problem is then to explain why Planck-suppressed operators should have such a flavor structure.}
Increasing the masses of the first and second generation squarks
does not save the situation because we can draw a diagram including only the
light sparticles required to preserve naturalness, as in figure~\ref{fig:protondecay}.

A simple solution is to impose either lepton- or baryon-number conservation on the theory. While $U(1)_B$ and $U(1)_L$ are anomalous, whereas $U(1)_{B-L}$ does not forbid $QQQ L$ (nor does R-parity), anomaly-free discrete subgroups $\mathbb{Z}_3^{(B)} \in U(1)_B$ and $\mathbb{Z}_3^{(L)} \in U(1)_L$ exist, due to the existence of three generations of matter in the standard model.\footnote{Discrete anomalies were first discussed in~\cite{Ibanez:1991hv}. However, of the ``anomalies'' discussed in that work, only the $G^2 \mathbb{Z}_n$ and $({\rm grav})^2 \mathbb{Z}_n$ anomalies (for $G$ nonabelian) are required to vanish for a conserved symmetry~\cite{Banks:1991xj}, relating to gauge and gravitational instantons, respectively.} We focus on the leptonic $\mathbb{Z}_3^{(L)}$, 
\begin{equation} \label{eqn:Z3Lcharges}
\begin{split}
\\[-2.5ex]
L \rightarrow e^{2\pi i/ 3} \, L, \qquad \bar{\nu} \rightarrow e^{-2\pi i/ 3} \, \bar{\nu}, \qquad \bar{e} \rightarrow e^{-2\pi i/ 3} \, \bar{e}\,,\\[1ex]
\end{split}
\end{equation}
which forbids $QQQ L$ as well as many other dangerous operators.
Majorana neutrino masses are forbidden by the $\mathbb{Z}_3^{(L)}$ symmetry, so we are forced to introduce Dirac masses for the neutrinos.\footnote{
Alternatively, we can explicitly break the lepton number symmetry by the Majorana neutrino masses of the see-saw mechanism. This case requires a careful study of the possible contributions to proton decay, as in e.g.~\cite{MFVSUSY}.} The tiny neutrino Yukawa coupling can be explained if $\bar{\nu}$ is strongly localized toward the UV brane \cite{Grossman:1999ra}.

The discrete lepton-number symmetry imposed above still allows some lepton-number violating couplings, such as the superpotential coupling $\bar{\nu}\bar{\nu}\bar{\nu}$ on both of the branes and $\frac{1}{\Lambda_{\rm IR}^3} (L H_u)^3$ on the IR brane, which could lead to proton decay in the presence of baryon number violation.
However, every proton decay diagram involving the coupling $\bar{\nu}\bar{\nu}\bar{\nu}$ is suppressed by the neutrino Yukawa coupling
and hence the decay width is strongly suppressed.
The contribution from the operator $(L H_u)^3$ to proton decay is also small enough to be ignored
because all lepton multiplets are localized toward the UV brane and dimension of the operator $(L H_u)^3$ is rather high.
In addition, proton decay to gravitinos is forbidden when the gravitino mass is larger than the nucleon mass.
Thus, there is no proton decay problem once $\mathbb{Z}_3^{(L)}$ is imposed, and neither R-parity nor baryon number conservation need to be imposed.

\subsection{Baryon number violation} \label{subsec:BNV}

While we imposed $\mathbb{Z}_3^{(L)}$ to suppress proton decay,
baryon number violation is allowed, which will lead to baryonic R-parity violation (RPV).
We will show that baryon number violation is suppressed by the wavefunction profiles of the quark multiplets as well as by the heavy squark masses, easily satisfying constraints  from $\Delta B = 2$ processes such as $n - \bar{n}$ oscillations and dinucleon decay. Conversely, the presence of R-parity violating couplings allows the lightest supersymmetric particle (LSP) to decay, relaxing experimental constraints on the stop mass and hence reducing tuning.

\begin{table}[t]
\renewcommand{\arraystretch}{1.3}
\begin{center}
\begin{tabular}[t]{c|ccc}
 & $sb$ & $bd$ & $ds$ \\
\hline
$u$ & $8 \times 10^{-6}$  & $2 \times 10^{-6}$ & $1 \times 10^{-6}$ \\
$c$ & $7 \times 10^{-4}$  & $2 \times 10^{-4}$ & $1 \times 10^{-4}$ \\
$t$ & $3 \times 10^{-3}$  & $1 \times 10^{-3}$ & $6 \times 10^{-4}$ \\
\end{tabular}
\vspace{0.3cm}
\caption{Estimate of the effective RPV couplings originating on the IR brane for $\tan \beta = 3$ and $\zeta_{Q_3} = 1$ using~(\ref{eqn:lowscalequark}). These are suppressed by the wavefunction profiles~(\ref{wffactor}).}
\label{RPVnumerical}
\end{center}
\renewcommand{\arraystretch}{1}
\end{table}

The baryonic R-parity violating couplings on the IR brane, consistent with all the symmetries in the theory, are given by
\begin{equation}
\begin{split}
\\[-2.5ex]
S_{\rm RPV, \, IR} = \int d^5x \, \delta(y-\pi R) \, e^{-3k\pi R} \left( \int d^2 \theta \, \frac{1}{2} \tilde{\lambda}''^{\, ijk}_{\rm IR} \bar{u}_i \bar{d}_j \bar{d}_k + {\rm h.c.} \right), \\[0.5ex]
\end{split}
\end{equation}
where the coupling $\tilde{\lambda}''_{\rm IR}$ has mass dimension $-3/2$. (A factor of $1/2$ is included due to the antisymmetry of the operator on exchange of the two down-type quark multiplets.)
Using the wavefunction profile \eqref{profile} and integrating over $y$, we find the 4D effective superpotential,
\begin{equation}
\begin{split}
W^{4D}_{\rm RPV, \, IR} = \frac{1}{2} {\lambda}''^{\, ijk}_{\rm IR} \bar{u}_i \bar{d}_j \bar{d}_k,
\label{RPVcoupling} \\[1ex]
\end{split}
\end{equation}
where $\lambda''^{\, ijk}_{\rm IR} = \tilde{\lambda}''^{\, ijk}_{\rm IR} k^{3/2} \zeta_{\bar{u}_i} \zeta_{\bar{d}_j} \zeta_{\bar{d}_k}$ is dimensionless and $\zeta_{\Psi}$ is given by~\eqref{zetaexpression}.
Using~\eqref{wffactor} and assuming that $k^{3/2} \tilde{\lambda}''^{\, ijk}_{\rm IR}$ is $\mathcal{O}(1)$ with anarchic flavor structure, we estimate
\begin{equation}
\begin{split}
\\[-2ex]
&\lambda''^{\, usb}_{\rm IR} \simeq \frac{m_u m_s m_b}{\lambda^5 \zeta_{Q_3}^3 v^3 s_\beta c_\beta^{2}},
\qquad \lambda''^{\, ubd}_{\rm IR} \simeq \frac{m_u m_b m_d}{\lambda^6 \zeta_{Q_3}^3 v^3 s_\beta c_\beta^{2}},
\qquad \lambda''^{\, uds}_{\rm IR} \simeq \frac{m_u m_d m_s}{\lambda^8 \zeta_{Q_3}^3 v^3 s_\beta c_\beta^{2}}, \\[1.5ex]
&\lambda''^{\, csb}_{\rm IR} \simeq \frac{m_c m_s m_b}{\lambda^4 \zeta_{Q_3}^3 v^3 s_\beta c_\beta^{2}},
\qquad \lambda''^{\, cbd}_{\rm IR} \simeq \frac{m_c m_b m_d}{\lambda^5 \zeta_{Q_3}^3 v^3 s_\beta c_\beta^{2}},
\qquad \lambda''^{\, cds}_{\rm IR} \simeq \frac{m_c m_d m_s}{\lambda^7 \zeta_{Q_3}^3 v^3 s_\beta c_\beta^{2}}, \\[1.5ex]
&\lambda''^{\, tsb}_{\rm IR} \simeq \frac{m_t m_s m_b}{\lambda^2 \zeta_{Q_3}^3 v^3 s_\beta c_\beta^{2}},
\qquad \lambda''^{\, tbd}_{\rm IR} \simeq \frac{m_t m_b m_d}{\lambda^3 \zeta_{Q_3}^3 v^3 s_\beta c_\beta^{2}},
\qquad \lambda''^{\, tds}_{\rm IR} \simeq \frac{m_t m_d m_s}{\lambda^5 \zeta_{Q_3}^3 v^3 s_\beta c_\beta^{2}}, \label{RPVcouplingIR} \\[1.5ex]
\end{split}
\end{equation}
up to order-one factors.
Note that, apart from the entries $\lambda''^{\, ubd}_{\rm IR}, \lambda''^{\, uds}_{\rm IR}, \lambda''^{\, cds}_{\rm IR}$ and the overall normalization, the structure of these couplings is identical to that of MFV SUSY~\cite{MFVSUSY}. However, due to the presence of UV brane localized couplings and the decoupling of most of the squarks, this will not play a large role in our analysis.

We can numerically estimate the RPV couplings~(\ref{RPVcouplingIR}) using~(\ref{eqn:TeVquarkmass}). However, for the purpose of estimating the rates of baryon-number violating processes we must account for the RG enhancement of the six-quark $\Delta B = 2$ effective operators generated after integrating out the superpartners. As a crude approximation, we account for these effects by using
the following low-energy quark masses in place of~(\ref{eqn:TeVquarkmass}):
\begin{equation}
\begin{split} \label{eqn:lowscalequark}
\\[-2.5ex]
&m_u \sim 3 \, {\rm MeV}, \qquad 
m_c \sim 1.3 \, {\rm GeV}, \qquad m_t \sim 173 \, {\rm GeV}, \\[1ex]
&m_d \sim 6 \, {\rm MeV}, \qquad 
m_s \sim 100 \, {\rm MeV}, \qquad m_b \sim 4 \, {\rm GeV}, \\[1ex]
\end{split}
\end{equation}
as in~\cite{MFVSUSY}.
The numerical values of the resulting couplings are shown in table~\ref{RPVnumerical} for $\tan \beta = 3$ and $\zeta_{Q_3} = 1$.

\begin{table}[!t]
\renewcommand{\arraystretch}{1.3}
\begin{center}
\begin{tabular}[t]{c|ccc}
 & $sb$ & $bd$ & $ds$ \\
\hline
$u$ & $0.04$  & $0.05$ & $0.05$ \\
$c$ & $0.02$  & $0.03$ & $0.03$ \\
$t$ & $5 \times 10^{-16}$  & $6 \times 10^{-16}$ & $6 \times 10^{-16}$ \\
\end{tabular}
\vspace{0.3cm}
\caption{Estimate of the effective RPV couplings originating on the UV brane for $\tan \beta = 3$ and $\zeta_{Q_3} = 1$.}
\label{RPVnumericalUV}
\end{center}
\renewcommand{\arraystretch}{1}
\end{table}

If no additional symmetries are imposed, then RPV couplings can also appear on the UV brane.\footnote{RPV couplings cannot appear in the bulk due to the presence of extended supersymmetry.}
The baryonic R-parity violating couplings on the UV brane, consistent with all the symmetries in the theory, are given by
\begin{equation}
\begin{split}
\\[-2.5ex]
S_{\rm RPV, \, UV} = \int d^5x \, \delta(y)  \left( \int d^2 \theta \, \frac{1}{2} \tilde{\lambda}''^{\, ijk}_{\rm UV} \bar{u}_i \bar{d}_j \bar{d}_k + {\rm h.c.} \right)\,. \\[1ex]
\end{split}
\end{equation}
As above, we find the effective superpotential
\begin{equation}
\begin{split}
\\[-2.5ex]
W^{4D}_{\rm RPV, \, UV} = \frac{1}{2} {\lambda}''^{\, ijk}_{\rm UV} \bar{u}_i \bar{d}_j \bar{d}_k,
\label{UVRPVcoupling} \\[1ex]
\end{split}
\end{equation}
where $\lambda''^{\, ijk}_{\rm UV} = \tilde{\lambda}''^{\, ijk}_{\rm UV} k^{3/2} \eta_{\bar{u}_i} \eta_{\bar{d}_j} \eta_{\bar{d}_k}$ is dimensionless and $\eta_{\Psi}$, given by~(\ref{etaexpression1}), is exponentially suppressed for the top multiplet.
Numerical estimates of these couplings are shown in table~\ref{RPVnumericalUV} for $\tan \beta = 3$ and $\zeta_{Q_3} = 1$. While the couplings involving the top/stop are negligible, the remaining couplings are large. Nonetheless, these couplings are not dangerous due to the large squark masses for the other flavors. A squark mass of $5 \times 10^4$ TeV leads to an additional suppression of $10^{-10}$ versus a squark mass of e.g.\ $500$ GeV, which makes these couplings comparable to the smallest couplings encountered in the MFV SUSY scenario, c.f.~\cite{MFVSUSY}.

\subsection{$n-\bar{n}$ oscillations}



The RPV couplings~\eqref{RPVcoupling} and~\eqref{UVRPVcoupling} lead to baryon number violating processes which are constrained by low-energy measurements. While some aspects of our model are similar to MFV SUSY, the RPV couplings involving the top quark multiplet are somewhat larger in our case, whereas most of the squarks are decoupled, necessitating a reanalysis of possible contributions to $\Delta B = 2 $ processes.

\begin{figure}[t]
 \begin{center}
\subfigure[\label{sfig:nnbarIRamp}]{\includegraphics[clip, width=6cm]{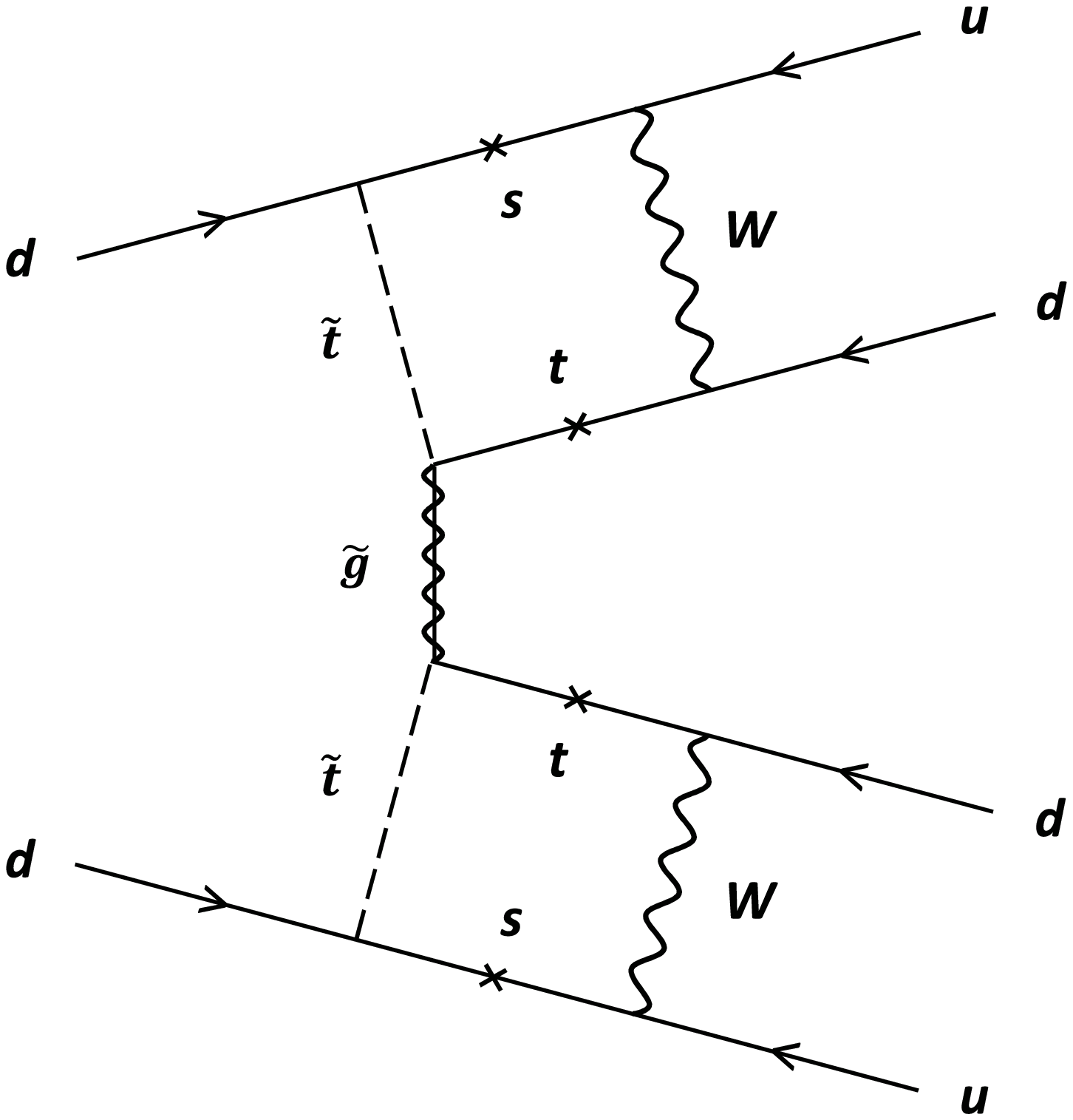}} \hspace{8mm}
\subfigure[\label{sfig:nnbarUVamp}]{\raisebox{9mm}{\includegraphics[clip, width=6cm]{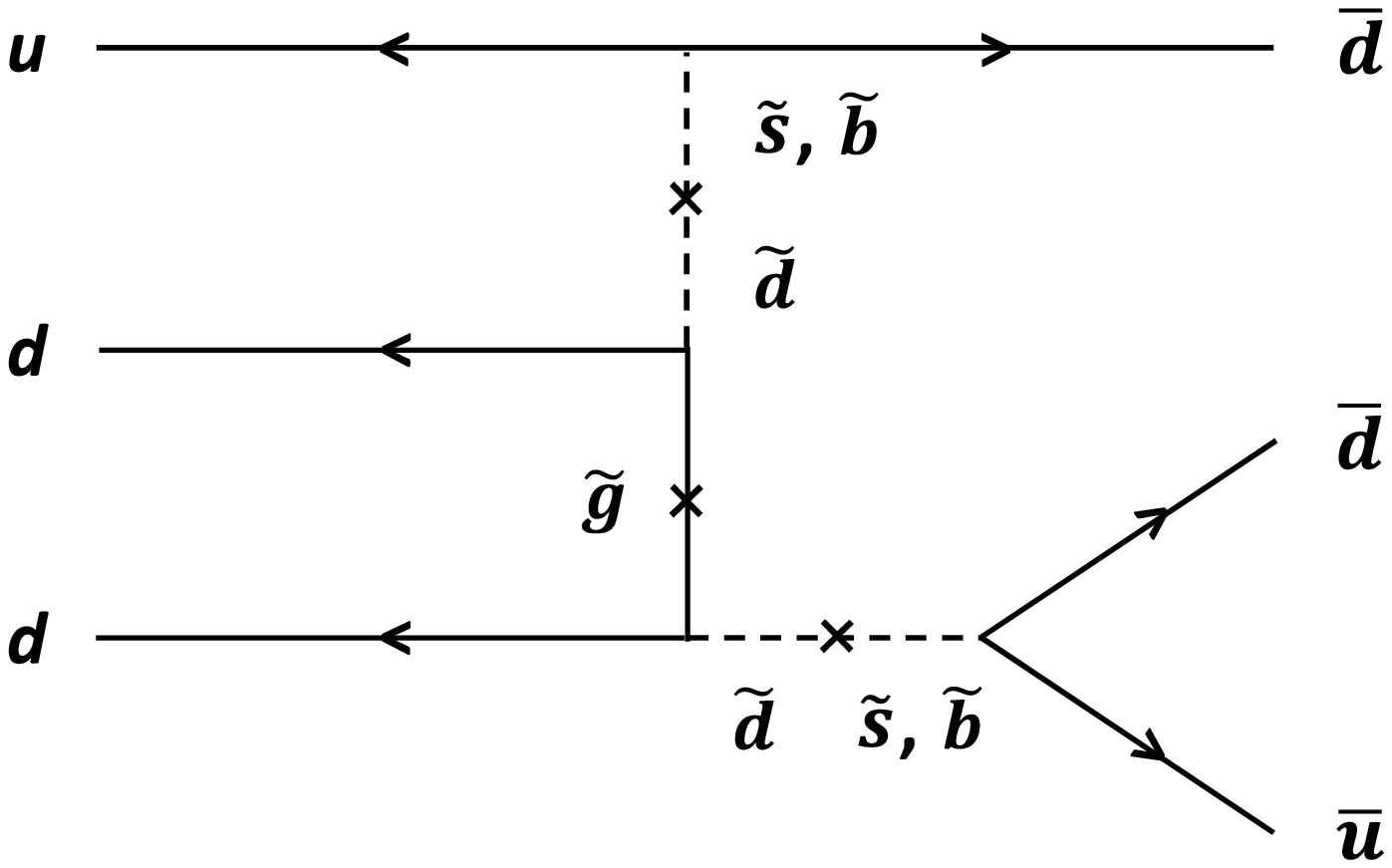}}}
    \caption{The leading contribution to $n - \bar{n}$ oscillations involving
only light superpartners such as stops and gauginos~\subref{sfig:nnbarIRamp}, and incorporating 
 heavy scalar superpartners but sizable RPV couplings~\subref{sfig:nnbarUVamp}.\label{fig:nnbar}}
  \end{center}
\end{figure}

The experimental lower bound on the $n - \bar{n}$ oscillation time is 
\cite{Abe:2011ky}
\begin{equation}
\begin{split}
\tau_{n-\bar{n}} \geq 2.44 \times 10^8 \, {\rm s}. \label{nnbarconst}
\end{split}
\end{equation}
We first consider diagrams involving only the light fields with masses less than $\Lambda_{\rm IR}$. The leading contribution to $n - \bar{n}$ oscillations in this case is the two-loop diagram shown in figure~\ref{sfig:nnbarIRamp}. Tree level and one loop diagrams are not possible due to the absence of tree-level FCNCs once the other squarks are decoupled.
We roughly estimate the amplitude of the diagram as
\begin{equation}
\begin{split}
\\[-2.5ex]
\mathcal{M}_{n- \bar{n}} \, \sim \, 4\pi \alpha_3 \left(\lambda''^{\, tds}_{\rm IR} \right)^2 \left( \frac{\alpha_2}{4\pi} \right)^2 \, \tilde{\Lambda}
\left( \frac{m_s \tilde{\Lambda}^2}{m_t m_{\tilde{t}}^2} \right)^2
\left( \frac{\tilde{\Lambda}}{m_{\tilde{g}}} \right), \\[1ex]
\end{split}
\end{equation}
where $\alpha_3 \simeq 0.1$ and $\alpha_2 \simeq 0.03$ are the $SU(3)_C$ and $SU(2)_W$ gauge couplings, respectively, and $\lambda''^{\, tds}_{\rm IR}$ is given by~\eqref{RPVcouplingIR}.
The factor $\tilde{\Lambda}^{6}$ comes from the hadronic matrix element, $\tilde{\Lambda} \sim \Lambda_{\rm QCD} \sim 250 \, \rm MeV$.
The $n - \bar{n}$ oscillation time is then $\tau_{n - \bar{n}} \sim \mathcal{M}_{n- \bar{n}}^{-1}$.
With reasonable values of $\tan \beta$, the stop mass and the gluino mass, the oscillation time is much longer than the bound~\eqref{nnbarconst}:
\begin{equation}
\begin{split}
\\[-2.5ex]
\tau_{n - \bar{n}} \, \sim \, \left( 3 \times 10^{10} \, {\rm s} \right) \zeta_{Q_3}^6 \biggl( \frac{3}{\tan \beta} \biggr)^4
\biggl( \frac{m_{\tilde{g}}}{1.2 \, \rm TeV} \biggr) \biggl( \frac{m_{\tilde{t}}}{300 \, \rm GeV} \biggr)^{4}. \\[1ex]
\end{split}
\end{equation}
Therefore, $n - \bar{n}$ oscillations coming from the diagram which involves
only light superpartners do not give a strict constraint on our model.

\begin{figure}[t]
 \begin{center}
   \subfigure[\label{sfig:nnbarUVcons}]{\includegraphics[clip, width=6cm]{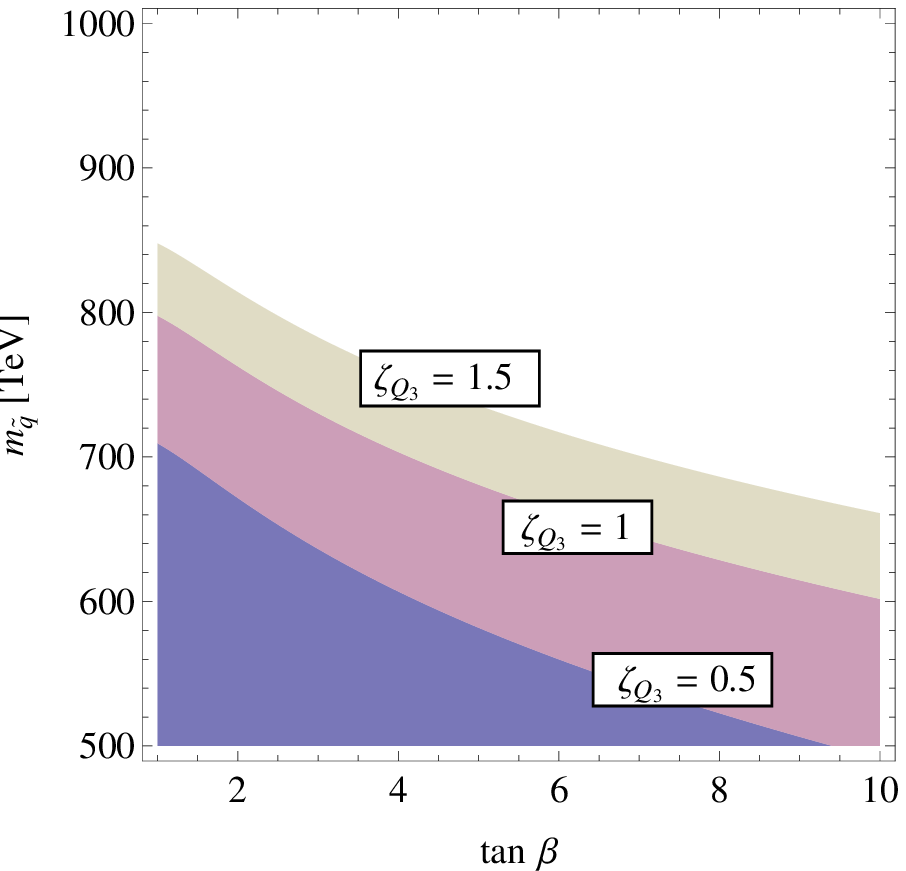}}\hspace{8mm}
    \subfigure[\label{sfig:dinucUVcons}]{\includegraphics[clip, width=6cm]{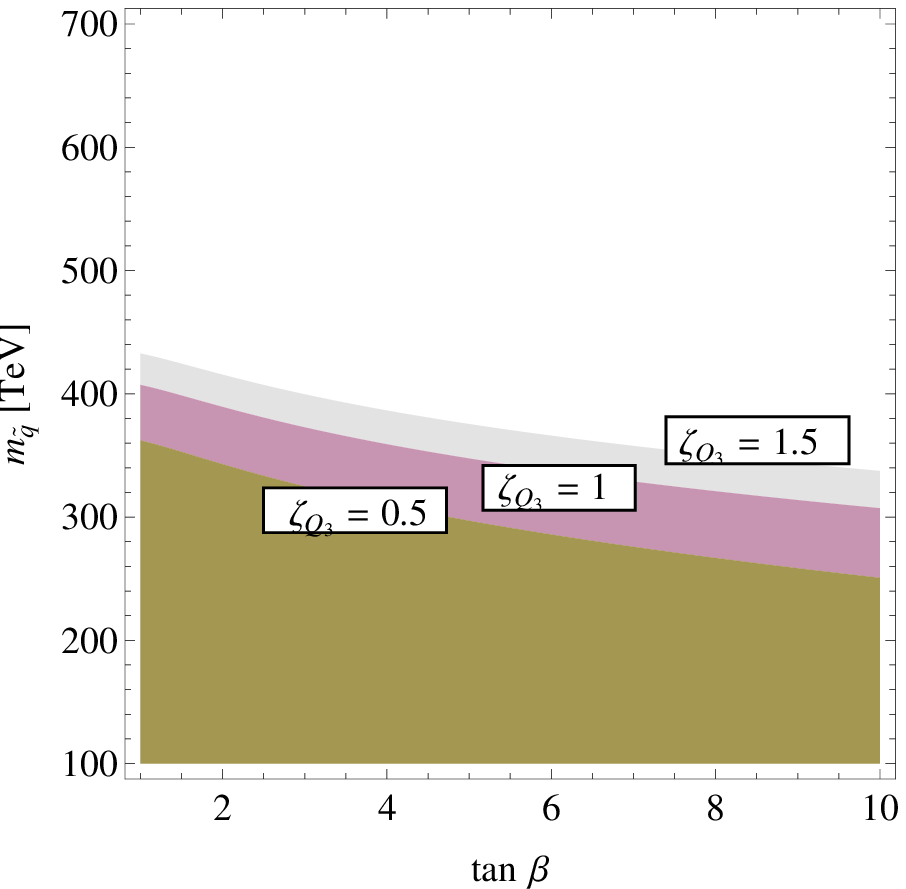}}
    \caption{Constraints on $\tan \beta$ and the heavy squark masses from $n - \bar{n}$ oscillations~\subref{sfig:nnbarUVcons} and
    dinucleon decay~\subref{sfig:dinucUVcons}.
The colored region is excluded. We assume $m_{\widetilde{g}} = 1.2 \, \rm TeV$ with the three colored regions corresponding to $\zeta_{Q_3} = 0.5, 1, 1.5$.}
    \label{fig:nnbarUV_const}
  \end{center}
\end{figure}

Although the scalar superpartners of the light quark multiplets are very heavy,
these multiplets feel sizable baryon number violation on the UV brane, and can in principle induce dangerous baryon number violation in the low-energy effective theory.
In figure~\ref{sfig:nnbarUVamp}, we show the leading tree-level contribution to $n - \bar{n}$ oscillations due to the exchange of heavy squarks.
We roughly estimate the amplitude of the diagram as
\begin{equation}
\begin{split}
\\[-2.5ex]
\mathcal{M}_{n- \bar{n}} \, \sim \, 4\pi \alpha_3 \left( \lambda''^{\, uds}_{\rm UV} \right)^2 \, \tilde{\Lambda}
\left( \frac{\tilde{\Lambda}}{m_{\tilde{g}}} \right) \left( \frac{\tilde{\Lambda}^2}{m_{\tilde{q}}^2} \right)^2, \\[1ex]
\end{split}
\end{equation}
where $m_{\tilde{q}}$ denotes the heavy squark mass, assumed to be of the same order for all the UV-brane-localized fields, and
the RPV coupling $\lambda''^{\, uds}_{\rm UV}$ is given in table~\ref{RPVnumericalUV}. Using this estimate and the experimental bound~(\ref{nnbarconst}), we place constraints on $\tan\beta$ and the heavy squark mass, $m_{\tilde{q}}$, as shown in figure~\ref{sfig:nnbarUVcons}. These constraints are automatically satisfied when the much stronger FCNC constraint $m_{\tilde{q}} \gsim 5\times 10^4$ TeV is imposed.

There are also potential contributions to the oscillation amplitude from the KK modes,
but these are strongly suppressed due to the small wavefunction overlap between the UV-brane localized light quarks and the IR-brane localized KK modes.

\subsection{Dinucleon decay}

\begin{figure}[t]
 \begin{center}
  \subfigure[\label{sfig:dinucIR}]{\includegraphics[clip, width=6cm]{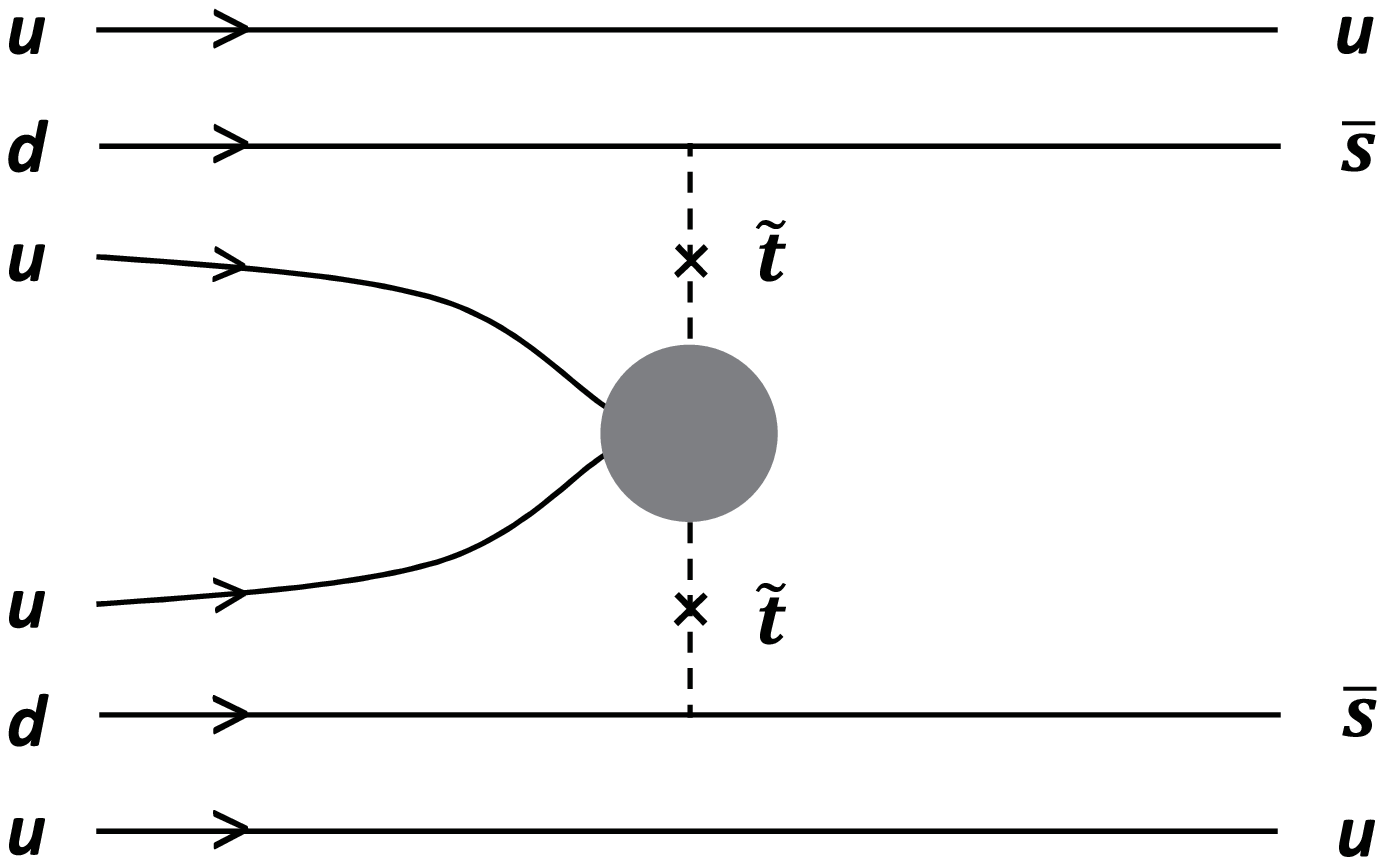}}\hspace{8mm}
    \subfigure[\label{sfig:dinucIRblob}]{\raisebox{6mm}{\includegraphics[clip, width=6cm]{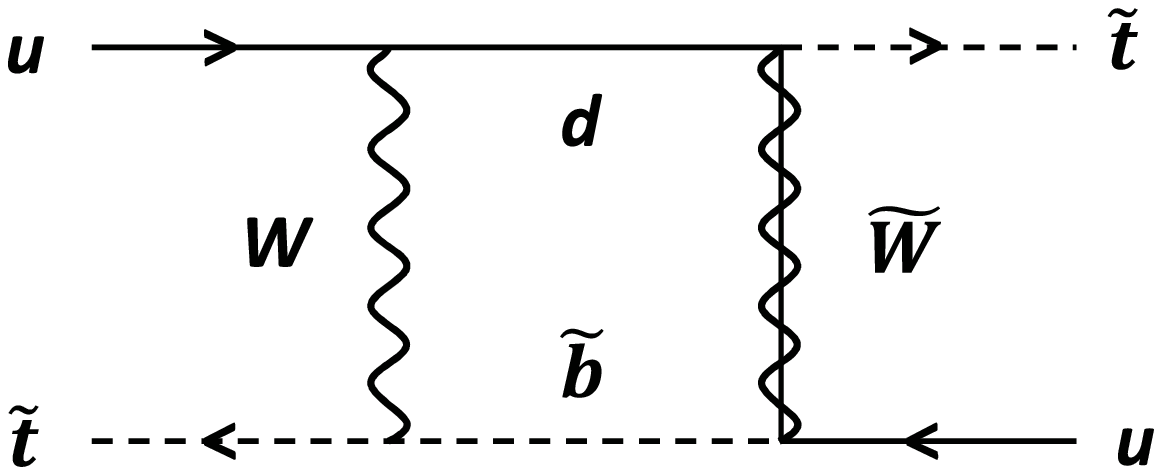}}}
    \caption{\subref{sfig:dinucIR}~The leading diagram contributing dinucleon decay, similar to~\cite{Goity:1994dq}.
The blob represents the box diagram~\subref{sfig:dinucIRblob}.\label{fig:dinucleon}}
\end{center}
\end{figure}

We next consider dinucleon decay.
The strongest limit typically comes from the lower bound on the partial lifetime for $pp \rightarrow K^+ K^+$ dinucleon decay~\cite{Litos:2010zra}
\begin{equation}
\begin{split}
\\[-2.5ex]
\tau_{pp \rightarrow K^+ K^+} \geq 1.7 \times 10^{32} \, {\rm yrs}. \label{dinucleonconst} \\[1.5ex]
\end{split}
\end{equation}
As above, we first consider diagrams involving only the light fields, in which case the leading contribution to dinucleon decay is the one-loop diagram shown in figure~\ref{fig:dinucleon}.

Following~\cite{Goity:1994dq}, we estimate the width for this diagram
\begin{equation}
\begin{split}
\\[-2.5ex]
\Gamma_{pp \rightarrow K^+ K^+} \, \sim \, \rho_N \, \frac{128 \pi \, \alpha_2^4 \left( \lambda''^{\, tds}_{\rm IR} \right)^4 \tilde{\Lambda}^{10}}{m_N^2 \, m_{\widetilde{W}}^2 \, m_{\tilde{t}}^8}
\left( \frac{\lambda^6  m_b^2}{4\pi m_{\tilde{b}}^2} \right)^2, \\[1ex]
\end{split}
\end{equation}
where $m_N \simeq m_p$ is the nucleon mass, $\rho_N \sim 0.25 \, {\rm fm}^{-3}$ is the nucleon density, $m_{\widetilde{W}}$ is the Wino mass, and $\tilde{\Lambda} \sim 250 \, \rm MeV$ is the scale associated to the hadronic matrix element.
The factor in parenthesis accounts for the flavor suppression which arises in the loop due to the GIM mechanism. We estimate the lifetime as
\begin{equation}
\begin{split}
\\[-2.5ex]
\tau_{pp \rightarrow K^+ K^+} \, \sim \, \left( 4 \times 10^{39} \, {\rm yrs} \right) \zeta_{Q_3}^{12} \biggl( \frac{3}{\tan \beta} \biggr)^8
\biggl( \frac{m_{\tilde{W}}}{600 \, \rm GeV} \biggr)^2 \biggl( \frac{m_{\tilde{t}, \, \tilde{b}}}{300 \, \rm GeV} \biggr)^{12}, \\[1ex]
\end{split}
\end{equation}
The experimental constraint is easily satisfied.



\begin{figure}
  \begin{center}
  \subfigure[\label{sfig:dinucUVamp}]{\includegraphics[clip, width=7cm]{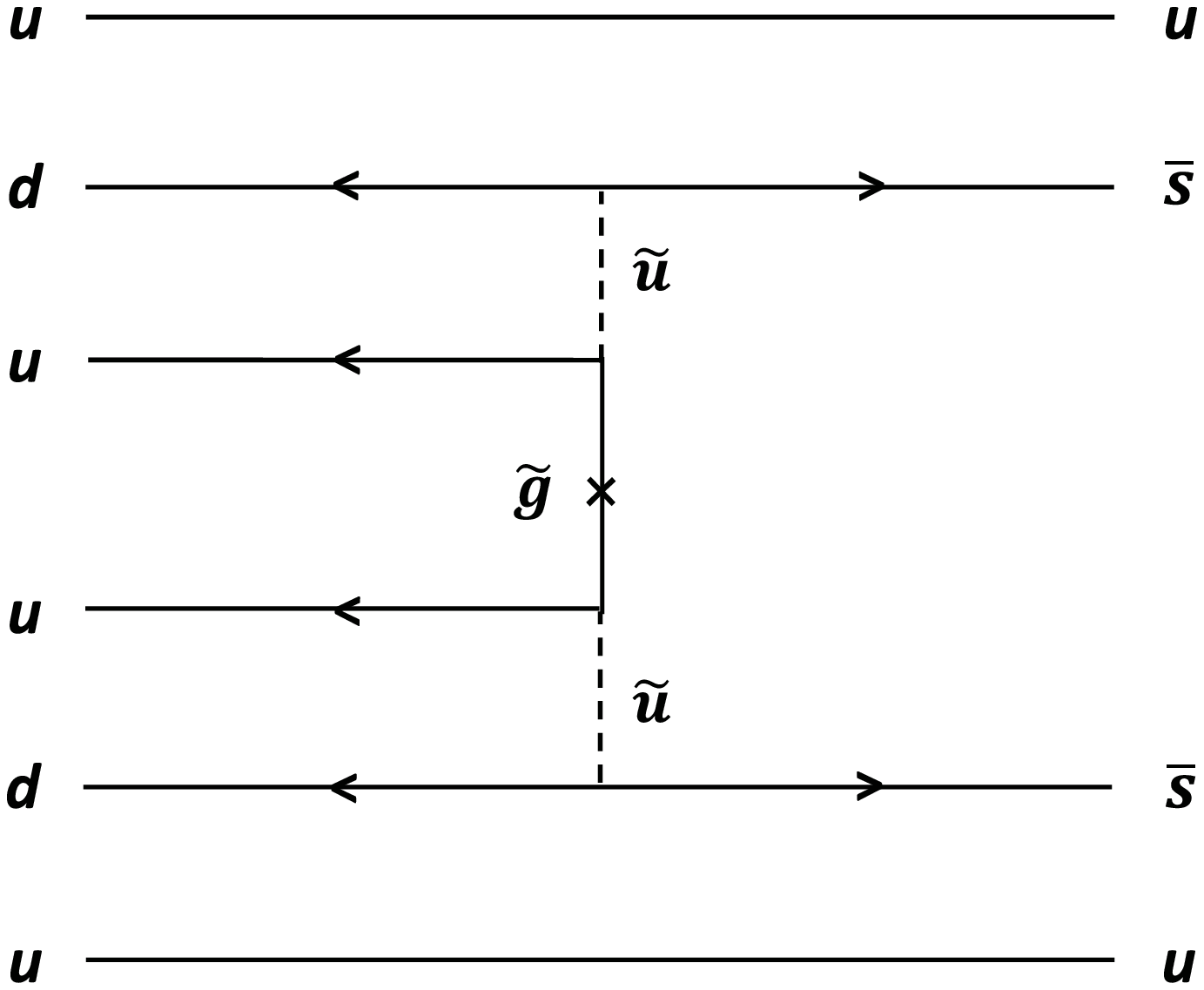}} \hspace{8mm}
  \subfigure[\label{sfig:LSPdecay}]{\raisebox{10mm}{\includegraphics[clip, width=6cm]{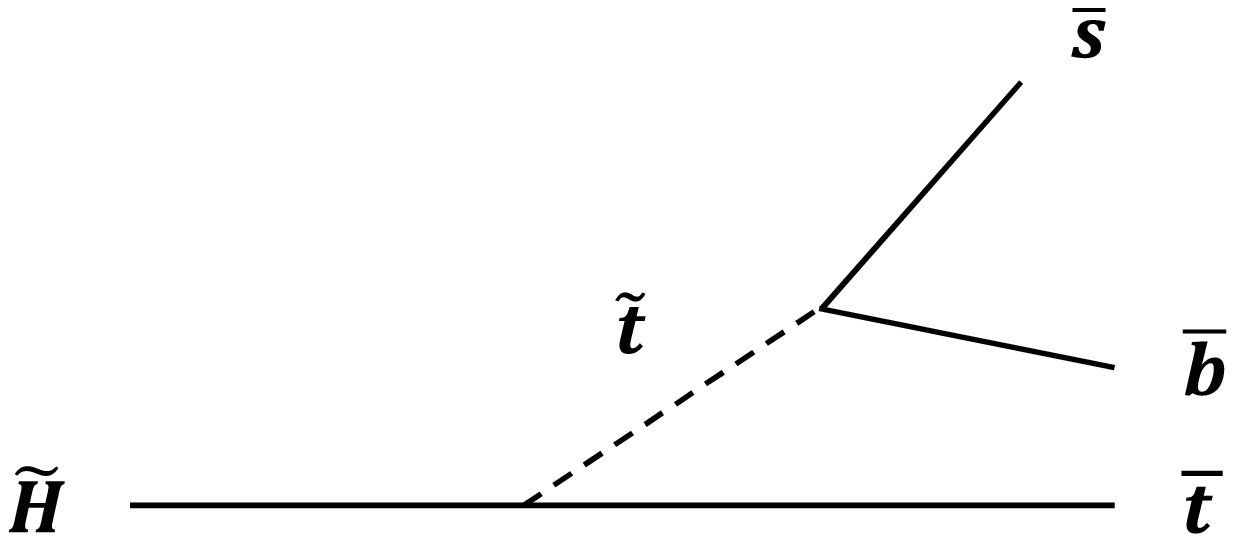}}}
    \caption{\subref{sfig:dinucUVamp}~The leading diagram contributing to dinucleon decay incorporating heavy squarks with sizable RPV couplings. \subref{sfig:LSPdecay}~The decay of a Higgsino-like LSP to standard model particles.}
  \end{center}
\end{figure}

Another possible contribution to dinucleon decay comes from diagrams which involve heavy squarks but sizable RPV couplings.
Figure~\ref{sfig:dinucUVamp} shows the leading diagram in this case.
As above, we estimate the $pp \rightarrow K^+ K^+$ width given by this diagram,
\begin{equation}
\begin{split}
\\[-2.5ex]
\Gamma_{pp \rightarrow K^+ K^+} \, \sim \, \rho_N \, \frac{128 \pi \, \alpha_3^2 \left( \lambda''^{\, uds}_{\rm UV} \right)^4 \tilde{\Lambda}^{10}}{m_N^2 \, m_{\widetilde{g}}^2 \, m_{\tilde{q}}^8}, \\[1ex]
\end{split}
\end{equation}
where the RPV coupling $ \lambda''^{\, uds}_{\rm UV}$ is the UV brane localized coupling.
Using this estimate and the experimental bound~(\ref{dinucleonconst}), we place constraints on $\tan\beta$ and the heavy squark mass, $m_{\tilde{q}}$, as shown in figure~\ref{sfig:dinucUVcons}. The constraints are weaker than those of $n - \bar{n}$ oscillations.
Therefore, the constraint from dinucleon decay is also satisfied in our scenario.

%

\subsection{LSP decay}




%


The lightest standard-model superpartner will be unstable due to the R-parity violating coupling~(\ref{RPVcoupling}). As discussed in~\S\ref{sec:SUSYbreaking}, the most likely LSP is the Higgsino with some chance of a stop LSP due to an accidental cancellation between positive and negative contributions to the stop mass. In either case, the LSP will decay promptly and without significant missing energy due to the relatively large R-parity violating couplings.

For instance, the leading decay channel for a neutral Higgsino LSP with $m_{\tilde{H}} > m_t$
is shown in figure~\ref{sfig:LSPdecay}. The width is approximately
\begin{equation}
\begin{split}
\\[-2.5ex]
\Gamma_{\widetilde{H}} \, \sim \, \frac{m_{\widetilde{H}}}{128 \pi^3} \, |\lambda_{tsb}|^2, \\[1ex]
\end{split}
\end{equation}
where $m_{\widetilde{H}}$ is the Higgsino mass. Using~(\ref{RPVcouplingIR}) and the quark masses~(\ref{eqn:TeVquarkmass}), we estimate a decay length of less than a micrometer, well beyond the capabilities of the LHC to detect.

If the neutralino is lighter than the top quark, it will decay via an off-shell top quark to a four or more body final state, and the width will incorporate additional phase-space suppression. However, the decay length is still too short to be observable. For a charged Higgsino LSP, a similar decay is possible, but with a bottom quark instead of a top quark in the final-state and hence no extra phase-space suppression. A stop LSP will decay via two jets. In either case, the decay is prompt.

\section{D-terms and unification} \label{sec:unification}

As explained in~\S\ref{sec:SUSYRS} and \S\ref{sec:SUSYbreaking}, in models of warped natural SUSY most of the squarks and sleptons are localized towards the UV brane, and obtain large masses well above the compactification scale via direct couplings to the SUSY breaking sector.
In this case, the hypercharge $D$-term can generate a dangerous correction to the Higgs soft masses~\cite{Sundrum:2009gv,Strassler:2003ht}.
After integrating out the heavy scalars, a large Fayet-Iliopoulos (FI) term can be generated in the effective theory
\begin{equation} \label{eqn:hyperchargeFI} 
\begin{split}
\mathcal{L}_{\rm FI} \sim \int d^4 \theta \, \frac{m_{\tilde{q}, \, \tilde{l}}^2}{16 \pi^2} \, g_Y V_Y, \\[1ex]
\end{split}
\end{equation}
where $m_{\tilde{q}, \, \tilde{l}}$ denotes the mass scale of the heavy scalars and $g_Y$ is the hypercharge gauge coupling. In the five-dimensional picture, the loop correction~(\ref{eqn:hyperchargeFI}) is generated near the UV brane, and hence -- unlike loop corrections on the IR brane -- it is not cut off at $\Lambda_{\rm IR}$. The resulting FI term propagates semi-classically through the bulk by inducing a vev for the adjoint chiral field $\Sigma$ in the $\mathcal{N}=2$ hypercharge vector multiplet~\cite{Gherghetta:2011wc}, and the IR-brane-localized scalars are not insulated from its effects.

This point is a little surprising, so we review it in the CFT picture as well~\cite{Sundrum:2009gv}. The five-dimensional bulk of the SUSY RS model corresponds to an approximate superconformal field theory (SCFT) which eventually confines at the compactification scale, corresponding to the appearance of the infrared brane. This SCFT talks to a SUSY breaking sector in the UV, as well a number of elementary multiplets with large SUSY breaking splittings (the UV-brane-localized matter fields). So long as the SUSY breaking deformations of the SCFT are irrelevant, or marginal ($\Delta \sim 4$) with small coefficients, the composite states (including the stop and the Higgs) will have small splittings, yielding the expected natural SUSY spectrum. 

The hypercharge vector multiplet $V_Y$ corresponds to an abelian conserved current $J^\mu_Y$ in the SCFT. Since it is conserved, $J^\mu_Y$ has scaling dimension $\Delta = 3$ exactly. However, supersymmetry mandates a scalar partner $D_Y$ for $J^\mu_Y$, whose scaling dimension is then $\Delta = 2$. Thus, the SCFT admits a relevant (supersymmetric) deformation:
\begin{equation}
\begin{split}
\\[-2.5ex]
\Delta \mathcal{L} = M_D^2 \, D_Y+ \cdots . \\[-2ex]
\label{eqn:DtermDefm}
\end{split}
\end{equation}
If this deformation is present in the UV theory, then the conformal phase will break down at the scale $M_D$. 
What happens at this scale will depend on the theory. In the presence of charged matter without a superpotential, the scalar component of the matter field will acquire a vev, Higgsing $U(1)_Y$. If this vev is prevented by a superpotential, then supersymmetry will be broken.

In either case, we must have $M_D \lsim \mathcal{O}(M_Z)$ to have a chance of reproducing the standard model with minimal tuning. However, if supersymmetry is broken in the UV theory then~(\ref{eqn:DtermDefm}) is induced by loop effects as in~(\ref{eqn:hyperchargeFI}), unless prevented by a symmetry. Cancelling the tree- and loop-level contributions
 requires substantial tuning in the UV theory, destroying naturalness.

\subsection{Traceless groups and exotics} \label{subsec:unification}

This problem arises because the $U(1)$ $D$-term is a gauge singlet. For a semi-simple gauge group $G$, no such relevant deformation exists, and even with $U(1)$ factors in $G$ the deformation can sometimes be forbidden by gauging an outer automorphism of $G$ under which $D_{U(1)}$ transforms nontrivially. The solution, then, is to embed the standard model in a semi-simple gauge group, or in a gauge group which admits an appropriate outer automorphism.

Groups with $U(1)$ factors but no singlet $D$-terms due to a gauged outer automorphism share some features in common with semi-simple groups. In particular, their representations must satisfy $\mathrm{Tr\ } T_{U(1)} = 0$, since the $U(1)$ generator $T_{U(1)}$ transforms in the same way as the $U(1)$ $D$-term, hence $\mathrm{Tr\ } T_{U(1)}$ is not a singlet and must vanish for a complete representation.\footnote{By contrast, an ordinary $U(1)$ gauge theory need only satisfy the anomaly cancellation condition $\mathrm{Tr\ } T_{U(1)} = 0$ for the fermion representations taken as a whole.\vspace{1mm}} We refer to groups without singlet $D$-terms (hence with traceless generators) as ``traceless'' groups for ease of discussion.

To ensure that the large soft masses on the UV brane do not generate dangerous $U(1)$ FI terms in the low energy effective theory, either the unbroken gauge group on the UV brane must be traceless, or else any $U(1)$ factors with singlet $D$-terms must neither couple to the light scalars nor mix with $U(1)$'s which do. Thus, the traceless component, $\hat{G}$, contains the standard model gauge group.
Moreover, $\hat{G}$ -- or a subgroup satisfying the same conditions -- must be unbroken in the bulk. Otherwise the bulk profiles of the standard model components of $\hat{G}$ irreps will be split, allowing an effective FI term to be generated.\footnote{A $\hat{G}$-breaking bulk vev is permissible so long as the profile is sufficiently IR-brane localized.}

Thus, we consider a traceless gauge group $\hat{G}$ which contains the standard model and is broken on the IR brane but preserved elsewhere. This configuration is reminiscent of Higgsless models of electroweak symmetry breaking~\cite{Higgsless}. As in these models, we find that the gauge bosons corresponding to the broken generators of $\hat{G}$ are generally about an order of magnitude lighter than the KK modes. Thus, these gauge bosons are a generic prediction of warped natural SUSY, and LHC constraints on them will provide an indirect constraint on the compactification scale, as discussed in~\S\ref{subsec:exoticCons}.

To embed hypercharge in a traceless group $\hat{G}$, it is necessary for the irreps to satisfy $\mathrm{Tr\ } T_Y = 0$. 
This implies that the standard model fermions must be embedded into larger representations. The $\hat{G}$ partners of each standard model fermion will consist of some combination of (i) other standard model fermions and (ii) ``exotics,'' i.e.\ new fermions not present in the standard model.

At first, exotics appear to be necessary; while several traceless models -- such as $SO(10)$ and its traceless subgroups 
-- unify the standard model fermions without exotics, the unified multiplets force the different standard model fermions to have the same bulk profiles, inconsistent with the observed Yukawa couplings and CKM matrix. Moreover, in some cases, such as for $SO(10)$ and $SU(5) \subset SO(10)$ models, unifying the standard model fermions into larger multiplets will induce proton decay mediated by the broken generators of the extended gauge group. Since the corresponding gauge bosons are light, this is disastrous.

The fermionic exotics may have the same standard model quantum numbers as the observed particles (as in orbifold GUTs
\cite{orbifoldGUT}, where the usual GUT multiplets are split by $\hat{G}$-breaking boundary conditions), or they may be different. In either case, 
unless they are sterile (neutral under the standard model gauge group), they must be sufficiently heavy to escape collider bounds, i.e.\ at least $\mathcal{O}(100\ \mathrm{GeV})$ to avoid LEP constraints, with some model dependence.

To achieve the requisite splitting between the standard model fields and the exotics, we could add $\hat{G}$-violating operators or boundary conditions on the IR brane.
However, the exotic partners of the first two generations of quarks and leptons are localized towards the UV brane, and the splitting which can be achieved by IR-brane-localized effects is consequently limited to $\Delta m \lsim \mathcal{O}(\zeta_\Psi k')$ or less, where $\zeta_\Psi$ is given by~(\ref{wffactor}). Thus, for $k' \sim 10$ TeV, at least some of the exotic partners of the first generation fermions will have masses $\mathcal{O}(10\ \mathrm{GeV})$ or less, inconsistent with LEP results.

This is a generic problem with warped natural SUSY models which (to our knowledge) has not previously been recognized. To solve it, we pursue a different approach: instead of splitting the multiplets, we find a way to split the effective Yukawa couplings of their components consistent with order-one couplings on the IR brane. To do so, we introduce multiple $\hat{G}$ multiplets in the bulk with a single zero mode between them due to the $\hat{G}$-invariant UV brane boundary conditions. A $U(1)$ symmetry imposed in the bulk and on the IR brane forces different bulk multiplets to couple to different $\hat{G}$-violating operators on the IR brane, so that the effective couplings for different components of the zero-mode will depend on different bulk mass parameters, allowing hierarchical couplings. This mechanism is explained in detail in~\S\ref{subsec:split} and applied to a left-right model in~\S\ref{subsec:left-right}.




\subsection{The $SU(5)$ model} \label{subsec:SU5}


We first discuss a simple $SU(5)$ model to illustrate the above points. We find that it is not viable due to the presence of light exotic fermions, an issue which will be addressed in the following subsections.

In the above discussion, we did not specify how the extended gauge group, $\hat{G}$, should be broken on the IR brane. Possibilities include $\hat{G}$-violating boundary conditions and spontaneous breaking via IR-brane-localized Higgs fields. In fact, these options are related, see e.g.~\cite{Csaki:2005vy}. For concreteness, we consider $SU(5)$ breaking by orbifold boundary conditions in the following discussion. We expect that other methods of $SU(5)$ breaking on the IR brane will have similar consequences.

To obtain different boundary conditions on the IR and UV branes, we start with a circle $y \cong y + 4 \pi R$ of twice the usual radius, and construct a $\mathbb{Z}_2 \times \mathbb{Z}_2'$ orbifold with the identifications $y \to -y$ and $y \to 2 \pi R - y$ under $\mathbb{Z}_2$ and $\mathbb{Z}_2'$, respectively. The orbifold action on the gauge fields take the form:
\begin{equation} \label{eqn:SUorbGauge}
\begin{split}
\\[-2.5ex]
\left( 
\begin{array}{cc}
V \\
\Sigma
\end{array}
\right) \to \left( 
\begin{array}{cc}
P V P^{\dag} \\
- P \Sigma P^{\dag}
\end{array}
\right) , \quad
\left( 
\begin{array}{cc}
V \\
\Sigma
\end{array}
\right) \to \left( 
\begin{array}{cc}
P' V P'^{\dag} \\
- P' \Sigma P'^{\dag}
\end{array}
\right) , \\[1ex]
\end{split}
\end{equation}
where $V$ and $\Sigma$ denote the vector and chiral components of the $\mathcal{N}=2$ bulk vector multiplet, and $P$ and $P'$ are $SU(5)$ matrices which encode the action of $\mathbb{Z}_2$ and $\mathbb{Z}_2'$, respectively, satisfying $P^2 = (P')^2 = 1$. We take $P=\mathrm{diag}(1,1,1,1,1)$ and $P'=\mathrm{diag}(1,1,1,-1,-1)$, so that $SU(5)$ is unbroken on the UV brane (the $\mathbb{Z}_2$ fixed point $y=0$) and is broken to $SU(3)\times SU(2)\times U(1)$ on the IR brane (the $\mathbb{Z}_2'$ fixed point $y = \pi R$). In the dual CFT picture, the $SU(5)$ symmetry is a weakly gauged flavor symmetry of the bulk CFT which is spontaneously broken at the confinement scale, much like chiral symmetry breaking in QCD.

To reproduce the correct gauge couplings in the low energy effective theory, we introduce IR-brane-localized kinetic terms for the standard model gauge fields:
\begin{equation}
\begin{split}
\\[-2.5ex]
S_{\rm IR} = \int d^5 x \, \delta (y - \pi R) \left\{ \sum_i  \frac{1}{4 \tilde{g}_i^2} \int d^2 \theta \, {\rm Tr} \, {W_i}^\alpha {W_i}_\alpha + {\rm h.c.} \right\}, \\[1ex]
\end{split}
\end{equation}
where $i=1,2,3$ labels the standard model gauge groups. By choosing the coefficients $1/ \tilde{g}_i^2$ appropriately, the difference between the observed gauge couplings $\alpha_1,\alpha_2,\alpha_3$ can be accommodated.

As implied above, the gauge sector has an interesting mass spectrum, given by the solutions of the equations~\cite{Gherghetta:2010cj}
\begin{equation}
\begin{split}
\\[-2.5ex]
\frac{J_0(m/k)}{Y_0(m/k)} &= \frac{J_0(m/k')}{Y_0(m/k')} \qquad \quad \text{for the NN gauge bosons}, \\[1.5ex]
\frac{J_0(m/k)}{Y_0(m/k)} &= \frac{J_1(m/k')}{Y_1(m/k')} \qquad \quad \text{for the ND gauge bosons}, \\[1.5ex]
\end{split}
\end{equation}
where $J_\alpha$, $Y_{\alpha}$ are Bessel functions of order $\alpha$ and NN, ND, DN, or DD denotes (in sequence) the UV and IR brane boundary conditions, which are either Neumann (N) or Dirichlet (D). The massive NN gauge bosons are standard model KK modes, whereas the ND gauge bosons correspond to the $SU(5)$ generators which are broken by Dirichlet boundary conditions on the IR brane.
These equations can be solved approximately in the large-volume limit, $\pi k R \gg 1$, where $- \frac{J_0(m/k)}{Y_0(m/k)} \approx \frac{1}{2 k R} \ll 1$. In this limit the mass spectrum of NN (ND) gauge bosons is approximately $m_n \approx u_{0,n} k'$ ($m_n \approx u_{1,n} k'$)
 where $u_{\alpha,n}$ denotes the $n$th zero of the order-$\alpha$ Bessel function $J_\alpha$. To a reasonable approximation, $u_{\alpha,n} \approx \left(n + \frac{2 \alpha-1}{4}\right) \pi$, so that $m_n \approx \left(n - \frac{1}{4} \right) \pi k'$ for the standard model KK gauge bosons and $m_n \approx \left(n + \frac{1}{4} \right) \pi k'$ for the KK modes of the broken generators.

However, in the latter case an additional solution exists in the regime $m \ll k'$, where $-\frac{J_1(m/k')}{Y_1(m/k')} \approx \frac{\pi (m/k')^2}{4}$. The mass is suppressed by the square root of the volume:
\begin{equation}
\begin{split}
\\[-2.5ex]
m_0 \simeq \sqrt{\frac{2}{\pi k R}} \, k' . \label{XYmass} \\[1.5ex]
\end{split}
\end{equation}
Unlike the KK modes, these modes have approximately flat profiles away from the IR brane, and are more closely analogous to the zero modes of the unbroken generators. They correspond to the $X,Y$ gauge bosons in an ordinary $SU(5)$ GUT, and can mediate rapid proton decay, depending on the details of the matter sector (to be discussed below).

The masses of the $X,Y$ gauge bosons are suppressed relative to those of the SM KK modes by $\frac{4}{3 \pi} \sqrt{\frac{2}{\pi k R}} \simeq \frac{1}{10}$. Thus, with a low compactification scale required by naturalness, these exotic gauge bosons are within reach of the LHC experiments, and searches for them can indirectly constrain the compactification scale.

\begin{figure}[t]
\begin{center}
\subfigure[\label{sfig:CFTblob}]{\raisebox{8mm}{\includegraphics[clip, width=6.5cm]{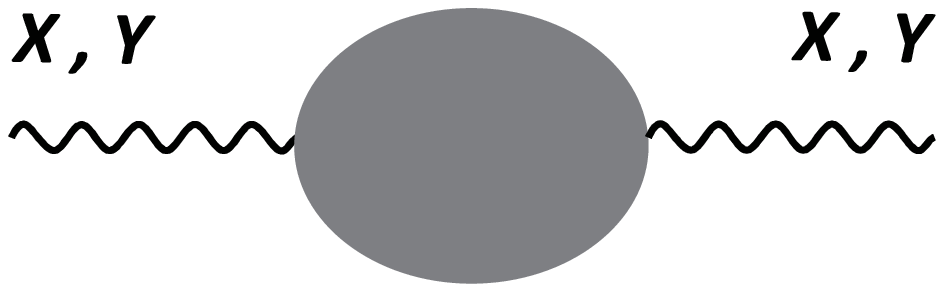}}} \hspace{8mm}
\subfigure[\label{sfig:lightProfile}]{\includegraphics[clip, width=6cm]{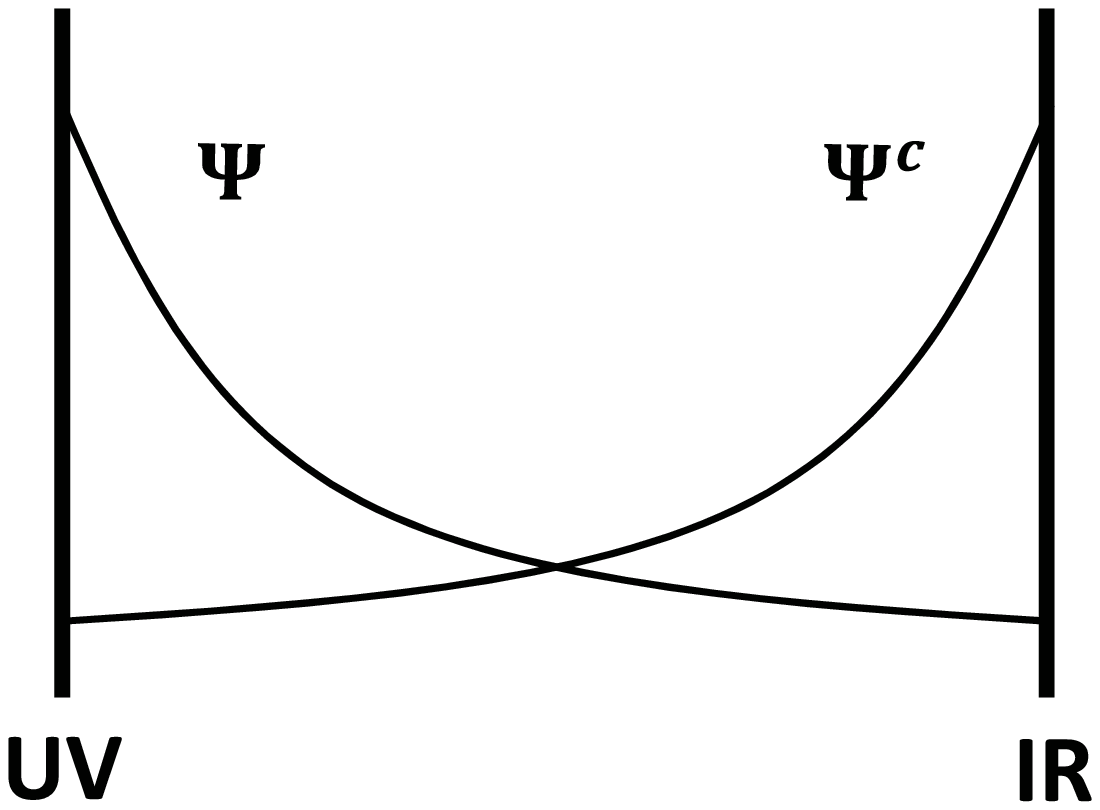}} 
    \caption{\subref{sfig:CFTblob}~A diagram contributing to the $X,Y$ gauge boson masses as well as renormalizing the $SU(5)$ gauge coupling. The blob denotes loops of CFT particles. \subref{sfig:lightProfile}~A schematic picture of the quasi-zero-mode $\Psi$ and $\Psi^c$ profiles with ND (DN) boundary conditions for $\Psi$ ($\Psi^c$) with $c_\Psi \gg 1/2$.}
    \label{fig:lightmode}
    
  \end{center}
\end{figure}

The appearance of these light modes can be understood in the dual CFT picture as follows. The masses of the $X,Y$ gauge bosons are generated by radiative corrections from the CFT particle states, as in figure~\ref{sfig:CFTblob}. The same diagram also induces logarithmic divergences in the $SU(5)$ theory, so we obtain the renormalized Lagrangian (c.f.~\cite{Burdman:2003ya})
\begin{equation}
\begin{split}
\\[-2.5ex]
\mathcal{L}_{SU(5)} \sim -\frac{1}{4} \left\{ \frac{1}{g_{\rm UV}^2}+\frac{N_{\rm CFT}}{16 \pi^2} \log \left( \frac{M_{\rm pl}}{\Lambda_{\rm IR}} \right) \right\}  \sum_a (F_{\mu \nu}^a)^2 + \frac{1}{2} \frac{N_{\rm CFT}}{16 \pi^2} \, \Lambda_{\rm IR}^2  \sum_{\alpha} (A_{\mu}^{\alpha})^2 \, , \\[1ex]
\end{split}
\end{equation}
where $g_{\rm UV}$ is the bare $SU(5)$ coupling at the Planck scale and $\alpha$ ($a$) indexes broken (all) generators of $SU(5)$. Neglecting $g_{\rm UV}$ (which is dual to a UV-brane-localized kinetic term for the gauge field), we canonically normalize to obtain the $X,Y$ gauge boson mass
\begin{equation}
\begin{split}
\\[-2.5ex]
m_0^2 \sim \frac{\Lambda_{\rm IR}^2}{\log \left( M_{\rm pl}/\Lambda_{\rm IR} \right)} \,, \\[1ex]
\end{split}
\end{equation}
in qualitative agreement with~(\ref{XYmass}), as $\log \left( M_{\rm pl}/\Lambda_{\rm IR} \right) \simeq \pi k R$.

We now introduce bulk hypermultiplets into the model. The orbifold action on the hypermultiplets is specified by the same matrices $P$, $P'$ as in~(\ref{eqn:SUorbGauge}), up to an overall choice of sign for each $\mathbb{Z}_2$ factor. We choose
\begin{equation}
\begin{split}
\\[-2.5ex]
&\left( 
\begin{array}{cc}
\Psi \\
\Psi^c
\end{array}
\right) \to \left( 
\begin{array}{cc}
P^{\dag} \Psi \\
- P^{\dag} \Psi^c
\end{array}
\right) , \quad
\left( 
\begin{array}{cc}
\Psi \\
\Psi^c
\end{array}
\right) \to \left( 
\begin{array}{cc}
\pm {P'}^{\dag} \Psi \\
\mp {P'}^{\dag} \Psi^c
\end{array}
\right) , \\[1.5ex]
&\left( 
\begin{array}{cc}
\Phi \\
\Phi^c
\end{array}
\right) \to \left( 
\begin{array}{cc}
P \Phi P \\
- P \Phi^c P
\end{array}
\right) , \quad
\left( 
\begin{array}{cc}
\Phi \\
\Phi^c
\end{array}
\right) \to \left( 
\begin{array}{cc}
\pm P' \Phi P' \\
\mp P' \Phi^c P'
\end{array}
\right) , \\[1.5ex]
\end{split}
\end{equation}
for $\Psi$ and $\Phi$ in the $\bar{\mathbf{5}}$ and $\mathbf{10}$ representations of $SU(5)$, respectively, in order to obtain zero modes in standard model representations. We label the resulting multiplets as ${\bf 10}_Q$, ${\bf 10}_{\bar{u}, \, \bar{e}}$, ${\bf \bar{5}}_{\bar{d}}$, and ${\bf \bar{5}}_{L}$, according to which zero modes they contain. By adding a bulk singlet $\bar{\nu}$ with NN boundary conditions, we recover a single generation of the standard model from five bulk multiplets. Although we are restricted to $\zeta_{\bar{u}} = \zeta_{\bar{e}}$ by this embedding, we can still reproduce the correct lepton masses by adjusting $\zeta_L$ and $\zeta_{\bar{\nu}}$ appropriately.

In addition to the zero modes $Q, \bar{u}, \bar{d}, L, \bar{e}, \bar{\nu}$, we have massive exotics, $Q', \bar{u}', \bar{d}', L', \bar{e}'$ and their vector-like partners, which fill out the $SU(5)$ multiplets. These exotics appear at or below the compactification scale, and interact with the standard model fermions via lepton- and baryon-number violating vertices with the $X,Y$ gauge bosons. Nonetheless, rapid proton decay can be avoided if there is no mixing between the exotics and the standard model fermions. In particular, we can assign $\mathbb{Z}_3^{(L)}$ charges to the multiplets as follows:
\begin{equation}
\begin{split}
\mathbf{10}_Q \to \omega_3 \mathbf{10}_Q, \;\;\; \mathbf{10}_{\bar{u},\, \bar{e}} \to \omega_3^{-1} \mathbf{10}_{\bar{u}, \, \bar{e}}, \;\;\; \bar{\mathbf{5}}_L \to \omega_3 \bar{\mathbf{5}}_L , \;\;\; \bar{\mathbf{5}}_{\bar{d}} \to \omega_3^{-1} \bar{\mathbf{5}}_{\bar{d}} \label{eqn:GUTZ3} \\[1ex]
\end{split}
\end{equation}
where $\omega_3 \equiv e^{2 \pi i/3}$. This agrees with the usual $\mathbb{Z}_3^{(L)}$ charges for the zero modes (\ref{eqn:Z3Lcharges}) combined with $Q \to \omega_3 Q,\, \bar{u} \to \omega_3^{-1} \bar{u},\, \bar{d} \to \omega_3^{-1} \bar{d}$, which is the action of the $\mathbb{Z}_3$ center of $SU(3)_C$; hence the two charge assignments are gauge-equivalent for the zero modes. However, $Q'$ now carries a different charge than $Q$, and likewise for the other exotics, so that mixing is forbidden. This is sufficient to stabilize the proton since the $SU(5)$ gauge interactions preserve $B-L$, and hence $X,Y$-mediated $\Delta B = 1$ processes are forbidden by $\mathbb{Z}_3^{(L)}$, which requires $\Delta L = 0 \mathrm{\ mod\ } 3$.

Unfortunately, as anticipated above this model has a fatal flaw due to the spectrum of exotics. As a warmup, we consider a single bulk hypermultiplet with chiral components $\Psi, \Psi^c$. Recall that the $\Psi$ zero mode (if it exists) is IR-brane localized for $c_\Psi \ll 1/2$ and UV-brane localized for $c_\Psi \gg 1/2$. Likewise, the zero mode for $\Psi^c$ (if it exists) is IR-brane localized for $c_\Psi \gg -1/2$ and UV-brane localized for $c_\Psi \ll -1/2$, since $\Psi \to \Psi^c$, $\Psi^c \to -\Psi$, $c_{\Psi}\to -c_{\Psi}$ is a symmetry of the theory. Suppose that $\Psi$ ($\Psi^c$) has NN (DD) boundary conditions, with $c_\Psi \gg 1/2$. In this case, there is a UV-brane localized $\Psi$ zero mode, whose support at the IR brane is exponentially suppressed. If we change the $\Psi$ boundary conditions to ND, the zero mode is lifted, but only a slight change to the profile is needed to satisfy the new boundary conditions. A similar argument shows that $\Psi^c$, whose boundary conditions are now DN, likewise has a zero-mode-like profile localized towards the IR brane, since a true IR-brane-localized $\Psi^c$ zero mode would exist after switching the UV-brane boundary condition to give DD (NN) for $\Psi$ ($\Psi^c$).

This suggests that for $c_\Psi \gg 1/2$ with ND (DN) boundary conditions for $\Psi$ ($\Psi^c$), both $\Psi$ and $\Psi^c$ have quasi-zero-modes with zero-mode-like profiles, which are respectively UV- and IR-brane localized. These modes cannot be massless, so it is natural to suppose that they pair up to get a mass. In this case, since their overlap is exponentially suppressed, their mass must also be exponentially suppressed, and these modes are much lighter than the compactification scale. This situation is depicted in figure~\ref{sfig:lightProfile}.

The above argument is heuristic, but can be verified by explicit computations. The mass spectrum for these boundary conditions is given by the solutions of (c.f.~\cite{Gherghetta:2010cj})
\begin{equation}
\begin{split}
\\[-2.5ex]
\frac{J_{c-1/2}(m/k)}{Y_{c-1/2}(m/k)} = \frac{J_{c+1/2}(m/k')}{Y_{c+1/2}(m/k')} \,. \\[1ex]
\end{split}
\end{equation}
For $c\gg 1/2$ and $m \ll k'$, we can approximate $-J_\alpha(u)/Y_\alpha(u)\approx \frac{\pi}{\Gamma(\alpha)\Gamma(\alpha+1)} \left(\frac{u}{2}\right)^{2 \alpha} + \mathcal{O}(u^{2 \alpha+2})$ for $\alpha > 0$ on both sides of the equation. We obtain the solution
\begin{equation}
\begin{split}
\\[-2.5ex]
m \simeq 2 \sqrt{c+ \frac{1}{2}} \, \zeta \, k' \\[1ex]
\end{split}
\end{equation}
where $\zeta$ is given by~(\ref{eqn:zeta}). In fact, this is an excellent approximation for all $c \gsim 1/2$, including the special case $c=1/2$ which matches~(\ref{XYmass}). Likewise, it is straightforward to verify that the corresponding $\Psi$ and $\Psi^c$ profiles closely approximate the zero mode with the same $c$. For $c\ll 1/2$, no light mode exists with these boundary conditions.

The appearance of a light mode for $c\gg 1/2$ ($c\ll -1/2$) with ND (DN) boundary conditions has been previously discussed in~\cite{Contino:2004vy}, where a CFT interpretation was given.
This is a disaster for the model outlined above, since e.g.\ the exotic $Q'_1$ will have a mass $M_{Q'_1} \sim 2 \zeta_{\bar{u}_1} k' \ll M_Z$, which is clearly ruled out.

\subsection{Split couplings without light exotics} \label{subsec:split}

We now show how
 the problem of light exotic fermions can be avoided. 
We begin with a toy model, consisting of two fermion multiplets $A$ and $B$, which unify into a single multiplet $\Psi = (A,B)$ in the bulk and on the UV brane. We couple $A$ and $B$ to operators on the IR brane:
\begin{equation} 
\begin{split}
\mathcal{L}_{\rm IR} = A \mathcal{O}_A + B \mathcal{O}_B + \ldots \,, \label{eqn:toyLIR} \\[1ex]
\end{split}
\end{equation}
which leads to the low-energy effective theory
\begin{equation}
\begin{split}
\\[-2.5ex]
\mathcal{L}_{\rm eff} = y_A \hat{A} \mathcal{O}_A + y_B \hat{B} \mathcal{O}_B + \ldots \,, \\[1ex]
\end{split}
\end{equation}
where $\hat{A}$ and $\hat{B}$ are canonically normalized massless fields. Our objective is to engineer hierarchical couplings in the effective theory $y_A \ll y_B \ll 1$ by controlling the profiles of the bulk fields with order-one changes in their bulk masses.

Motivated by orbifold-GUTs, we could introduce
two bulk multiplets, $\Psi_A = (A,B')$ and $\Psi_B = (A',B)$, with NN boundary conditions for the indicated zero mode and ND boundary conditions for the exotics $A'$ and $B'$. We then obtain $y_A \sim \zeta_A$ and $y_B \sim \zeta_B$, which can be adjusted independently using the bulk mass parameters $c_A$ and $c_B$. However, as argued above, the exotics $A'$ and $B'$ 
obtain exponentially suppressed masses $\mathcal{O}(\zeta_B k')$ and $\mathcal{O}(\zeta_A k')$, respectively. 
We must somehow lift these exotics to the compactification scale or above. 

We instead consider NN boundary conditions for $A'$ and $B'$, which will introduce additional zero modes. To compensate, we add a conjugate multiplet on the UV brane $\bar{\Psi}_{\rm UV} = (\bar{A}_{\rm UV}, \bar{B}_{\rm UV})$ and allow arbitrary $\mathcal{O}(k)$ mass terms $M_{\rm UV} \bar{\Psi}_{\rm UV} (s_\theta \Psi_A-c_\theta \Psi_B)$, where $\theta$ is an order-one mixing angle and $c_\theta \equiv \cos \theta$,  $s_\theta \equiv \sin \theta$. 
 As a result, only one linear combination $\hat{\Psi} = (\hat{A},\hat{B}) = c_\theta \Psi_A + s_\theta \Psi_B$ will be massless. 
 Accounting for the bulk profiles, (\ref{eqn:toyLIR}) gives the effective couplings $y_A \sim c_\theta \zeta_A$ and $y_B \sim s_\theta \zeta_B$, so we have removed the extra light states without sacrificing the splitting between $y_A$ and $y_B$.

However, due to the change in boundary conditions, additional couplings can now appear on the IR brane:
\begin{equation}
\begin{split}
\mathcal{L}_{\rm IR} = A' \mathcal{O}_A + B' \mathcal{O}_B + \ldots \,. \label{eqn:toyLIRadd} \\[1ex]
\end{split}
\end{equation}
In this case, we obtain $y_A \sim y_B \sim c_\theta \zeta_A + s_\theta \zeta_B$, which spoils the splitting. To avoid this, we introduce an abelian gauge symmetry (discrete or continuous) under which $\Psi_A$ and $\Psi_B$ carry different charges. We break the symmetry on the UV brane, allowing an arbitrary mixing angle $\theta$, but enforce it in the bulk and on the IR brane, aligning the bulk masses and forbidding~(\ref{eqn:toyLIRadd}).

This corresponds to an accidental flavor symmetry of the CFT. Even if we introduce a $U(1)$ symmetry, there is no $D$-term problem because the $D$-term obtains a large mass on the UV brane, precluding a $D$-term vev. In CFT language, the $U(1)$ flavor symmetry is not present in the UV theory, hence there is no conserved current $J_\mu$ and corresponding dimension-two $D$-term. This current and the corresponding relevant deformation only appear well into the CFT phase, hence the deformation is not excited by the large masses of fundamental fields in the UV theory.


A similar mechanism can generate $y_A \ll y_B \sim 1$ (relevant for the third generation) by taking $c_B \simeq -1/2$ and $c_A > -1/2$ such that $\eta_B / \eta_A \sim y_A$, where $\eta$ is given by~(\ref{eqn:etaexact}). In this case, the coupling of $\bar{\Psi}_{\rm UV}$ to $\Psi_A$ and $\Psi_B$ is suppressed by $\eta_A$ and $\eta_B$, respectively, and the massless combination is $\hat{\Psi} \sim \Psi_B + (\eta_B/ \eta_A) \Psi_A$, since $\eta_B \ll \eta_A$. With $\zeta_A, \zeta_B \sim \mathcal{O}(1)$, this reproduces the desired $y_A, y_B$. So long as $c_A \ge -1/2$, we have $\eta_A k \gsim k'$, and the massive linear combination of the $\Psi_A$ and $\Psi_B$ zero modes obtains a mass above the compactification scale, avoiding light exotics.


To avoid confusion between $\zeta_{A,B}$ (which is a function of $c_{A,B}$ only) and the profile of the true zero mode (which depends on the mixing angle induced by the UV brane boundary conditions) we denote the latter as $\hat{\zeta}_{A,B}$. In particular,
\begin{equation} 
\begin{split}
\\[-2.5ex]
\hat{\zeta}_A \equiv c_\psi \zeta_A\,,\;\;\; \hat{\zeta}_B \equiv s_\psi \zeta_B \,, \label{eqn:zetahat} \\[1 ex]
\end{split}
\end{equation}
where $\psi$ is the effective mixing angle, determined by $\tan \psi = \frac{\eta_A}{\eta_B} \tan \theta$.
The difference is pronounced for the third generation, where we can have $\hat{\zeta}_A \ll \hat{\zeta}_B$ while $\zeta_{A} \sim \zeta_B \sim 1$.


A few comments are in order. Firstly, adding $\bar{\Psi}_{\rm UV}$ on the UV brane with a large mass $M_{\rm UV} \bar{\Psi}_{\rm UV} (s_\theta \Psi_A-c_\theta \Psi_B)$ is similar to imposing off-diagonal boundary conditions on $\Psi_A$ and $\Psi_B$ on the UV brane:\footnote{These boundary conditions can be diagonalized by a unitary rotation between $\Psi_A$ and $\Psi_B$, but this creates off-diagonal bulk mass terms for $c_A \ne c_B$. We choose to work in the basis where the bulk masses are diagonal, leading to off-diagonal boundary conditions.}
\begin{equation}
\begin{split}
\\[-2.5ex]
\left(s_\theta \Psi_{A} - c_\theta \Psi_{B}\right)_{y=0} = 0\,, \;\;\; \left(c_\theta \Psi_{A}^c + s_\theta \Psi_{B}^c\right)_{y=0} = 0 \,.  \label{eqn:mixedBCs} \\[1ex]
\end{split}
\end{equation}
In particular, the two are exactly equivalent in the $M_{\rm UV} \to \infty$ limit, and qualitatively similar for $M_{\rm UV} \sim k$. Our discussion will not depend on $M_{\rm UV} \gsim k$, hence we treat these two possibilities as interchangeable.

Secondly, we can split a multiplet into more than two pieces using the same procedure. If $\Psi = (A_1, \ldots, A_n)$ is a unified bulk multiplet with $n$ standard model components whose couplings we wish to control individually, then we add $n$ copies $\Psi^i = (A^i_1, \ldots A^i_n)$, $i=1,\ldots,n$ and $n-1$ UV-brane localized $\bar{\Psi}$ multiplets. Allowing arbitrary mass terms on the UV brane, we obtain one light combination $\hat{\Psi} \sim \frac{1}{\sqrt{n}} \sum_i \Psi_i$. Imposing an abelian symmetry which forbids couplings to the ``off-diagonal'' components $A^i_j$, $i\ne j$, on the IR brane, we obtain independent couplings $y_i \sim \zeta_i/\sqrt{n}$ to the fields $\hat{A}_i$ on the IR brane. However, as before the mechanism relies on the existence of an appropriate symmetry to forbid the unwanted couplings. 

\subsection{A classification of traceless models}

The mechanism of~\S\ref{subsec:split} circumvents the problem of light exotics in traceless extensions of the standard model broken at the compactification scale, as required to forbid the relevant $D$-term deformation~(\ref{eqn:DtermDefm}) of the SCFT dual to the five-dimensional bulk. However, this mechanism does not involve a true splitting of the multiplets; the profiles of the different components remain identical unless mass terms or brane-localized kinetic terms are added on the IR brane, which in any case has little effect on the UV-brane localized fields. Instead, hierarchical couplings are generated by controlling which bulk fields can couple to which IR brane operators, where the degenerate zero modes are mix into several bulk multiplets.

As a consequence, UV-brane-localized fields will always appear in complete $\hat{G}$ multiplets in the effective theory. This precludes the use of models such as trinification ($SU(3)_C \times SU(3)_L \times SU(3)_R$) and the $E_6$ GUT, which rely on the introduction of additional charged and/or colored states at the unification scale. Inevitably, these states will appear as light exotics in the UV-brane-localized multiplets, ruling out these models.

Thus, $\hat{G}$ multiplets must consist of combinations of the standard model fermions with standard model singlets. Each multiplet must satisfy $\mathrm{Tr\ } T_Y = 0$, hence the possibilities are easily classified. Since $\mathrm{Tr\ } T_Y = 1$ for $Q$, $\bar{d}$, and $\bar{e}$, whereas $\mathrm{Tr\ } T_Y = -1$ for $L$ and $\mathrm{Tr\ } T_Y = -2$ for $\bar{u}$, we can form two traceless multiplets by combining $L$ with one of $\{Q, \bar{d}, \bar{e}\}$ and $\bar{u}$ with the remaining two, or we can combine all charged fermions into a single traceless multiplet.

The combinations $(L,\bar{d})$ and $(\bar{u},Q,\bar{e})$ occur in the $SU(5)$ model. There is a well known proton-decay problem due to the fact that the gauge interactions violate both baryon and lepton number (conserving $B-L$). This can be cured by splitting the multiplets, but the mechanism of~\S\ref{subsec:split} reintroduces the problem (since the multiplets are not really split).

The combinations $(L,Q)$ and $(\bar{u},\bar{d},\bar{e})$ occur in the Pati-Salam model, with gauge group $SU(4)\times SU(2)_L \times SU(2)_R$. For a low $SU(4)$ breaking scale, the broken generators of $SU(4)$ (which are leptoquarks) can mediate rare meson decays, such as $K_L \to \mu^{\pm} e^{\mp}$~\cite{Shanker:1981mj}. Although this violates both lepton and quark flavor, the diagonal flavor symmetries respected by the Pati-Salam gauge interactions are preserved, and there is no flavor suppression. Since the branching fraction is observed to be less than $4.7 \times 10^{-12}$~\cite{PDG} (the most common decays are $K_L \to \pi^{\pm} \ell^{\mp} \nu$), the $SU(4)$ breaking scale must be well above the weak scale, leading to excessive fine tuning. As before, this can be cured by splitting the multiplets, at the expense of light exotics which cannot be removed without reintroducing the problem.

The combinations $(L,\bar{e})$ and $(\bar{u},\bar{d},Q)$ occur in the minimal left-right model, discussed in the next section. In this case, neither proton decay nor rare meson decays are induced, and the model is viable with a relatively low compactification scale.

Finally, we can combine $(L,Q,\bar{u},\bar{d},\bar{e})$ into a single multiplet, as in the $SO(10)$ model and the left-right symmetric Pati-Salam model. However, these models have no advantages over their subgroups considered above, and both have problems with rare meson decays and/or proton decay.

Other traceless gauge groups with multiplets of the above types exist, but we know of no examples which avoid the proton decay and rare meson decay problems without introducing charged and/or colored exotics, apart from  models with the minimal left-right model as a subgroup.


\subsection{The left-right model} \label{subsec:left-right} \label{subsec:PatiSalam}

The minimal left-right model~\cite{leftright} is based on the ``3-2-2-1'' model, with an $SU(3)_C \times SU(2)_L \times SU(2)_R \times U(1)_{B-L}$ gauge group and matter in the $Q \equiv ({\bf 3}, {\bf 2}, {\bf 1})_{1/3}$, $U \equiv (\bar{\bf 3}, {\bf 1}, {\bf 2})_{-1/3}$, $L \equiv ({\bf 1}, {\bf 2}, {\bf 1})_{-1}$, and $E \equiv ({\bf 1}, {\bf 1}, {\bf 2})_{1}$ representations. 
Turning on a Higgs vev in the $({\bf 1}, {\bf 1}, {\bf 2})_{1}$ representation, 3-2-2-1 breaks to $SU(3)_C \times SU(2)_L \times U(1)_Y$, where hypercharge is generated by $T_Y = \frac{1}{2} T_{B-L} + \mathrm{diag}_{SU(2)_R} \left(\frac{1}{2},-\frac{1}{2}\right)$. The right-handed multiplets decompose $U \to (\bar{u}, \bar{d})$ and $E \to (\bar{e},\bar{\nu})$, reproducing the standard model without exotics. The 3-2-2-1 gauge interactions conserve $\mathbb{Z}_3 \in U(1)_{B}$ and $\mathbb{Z}_3 \in U(1)_{L}$ -- as in the standard model -- hence proton decay is not induced.

The matter content of the 3-2-2-1 model is symmetric under a \emph{left-right symmetry}, a $\mathbb{Z}_2$ outer automorphism which combines charge conjugation of $SU(3)_C$ and $U(1)_{B-L}$ with the exchange $SU(2)_L \leftrightarrow SU(2)_R$, so that $Q \leftrightarrow U$ and $L \leftrightarrow E$. Upon gauging the left-right symmetry, the multiplets further unify into irreps $\mathcal{Q} \equiv (Q,U)$ and $\mathcal{L} \equiv (L,E)$.
The $U(1)_{B-L}$ $D$-term is odd under the left-right symmetry, hence the relevant deformation~(\ref{eqn:DtermDefm}) is forbidden if the left-right symmetry is unbroken on the UV brane and in the bulk, which we assume henceforward. We refer to the 3-2-2-1 model with a gauged left-right symmetry as the (minimal) left-right model.

The left-right model cannot be broken to the standard model by orbifold boundary conditions, since the latter cannot reduce the rank of the gauge group~\cite{Csaki:2005vy}. Instead, we consider more general ``interval'' boundary conditions, which are either Neumann (N) or Dirichlet (D) for each field at each boundary, with opposite choices for the two $\mathcal{N}=1$ components of the $\mathcal{N}=2$ bulk vector- and hyper-multiplets. One can argue that, without access to a more fundamental description of the bulk theory such as an embedding into string theory, orbifold boundary conditions are not inherently more ``natural'' than interval boundary conditions~\cite{Csaki:2005vy}. Thus, we will not attempt to construct an orbifold model.\footnote{This can be done at the expense of introducing significantly more complicated physics on the IR brane.}

We choose NN boundary conditions for the standard model gauge bosons and ND boundary conditions for the additional gauge bosons, breaking $SU(2)_R \times U(1)_{B-L} \to U(1)_Y$ on the IR brane. As above, there will be light modes corresponding to the broken generators, leading to LHC constraints on the compactification scale, as discussed in~\S\ref{subsec:exoticCons}.


We now show that the correct Yukawa couplings can be reproduced with an unbroken left-right symmetry in the bulk and on the UV brane, using the techniques of~\S\ref{subsec:split}.
We focus on the quark sector and construct a 3-2-2-1 model, later incorporating the left-right symmetry. 
The simplest model which can accommodate the observed Yukawa couplings and CKM matrix consists of three bulk multiplets, $Q$, $U_{\bar{u}} = (\bar{u},\bar{d}')$ and $U_{\bar{d}} = (\bar{u}',\bar{d})$ with NN boundary conditions, coupled to a UV brane multiplet $\bar{U}$ to remove the extra right-handed zero modes. We introduce a $U(1)_X$ gauge symmetry broken by boundary conditions on the UV brane, under which $Q$ carries charge $q$, $U_{\bar{u}}$ ($U_{\bar{d}}$) carries charge $1-q$ ($-1-q$) and $H_d$ ($H_u$) carries charge $+1$ ($-1$). The allowed Yukawa couplings are
\begin{equation}
\begin{split}
\\[-2.5ex]
W_{\rm Yukawa} = Q \bar{u} H_u+Q \bar{d} H_d\,, \\[1ex] 
\end{split}
\end{equation}
for any $q$. We can adjust $\hat{\zeta}_Q$, $\hat{\zeta}_{\bar{u}}$, and $\hat{\zeta}_{\bar{d}}$ independently to reproduce the quark Yukawas and CKM matrix as usual, where $\hat{\zeta}_x$ is defined in~(\ref{eqn:zetahat}).

In order to incorporate the left-right symmetry, the bulk theory must contain complete left-right multiplets with equal bulk masses for the components. Since the profiles $\zeta_{Q_i}$, $\zeta_{\bar{u}_i}$ and $\zeta_{\bar{d}_i}$ differ for the three 3-2-2-1 multiplets, the most straightforward solution is to introduce left-right partners for each existing 3-2-2-1 multiplet to form $\mathcal{Q}_Q = (Q, U'')$, $\mathcal{Q}_{\bar{u}} = (Q',U_{\bar{u}})$ and $\mathcal{Q}_{\bar{d}} = (Q'',U_{\bar{d}})$. We now require two UV brane multiplets $\bar{\mathcal{Q}}_{1,2}$ to remove the excess zero modes. We assign $U(1)_X$ charges $p-q$ to $U''$ and $q+r$ ($q+s$) to $Q'$ ($Q''$), with $p\notin \{\pm1,\pm 1-r,\pm1-s\}$, $r,s\notin \{0,\pm 2\}$ to ensure that no additional Yukawa couplings are generated. Since the $U(1)_X$ charge assignments are not left-right invariant, we introduce an additional bulk gauge symmetry $U(1)_X'$ which is the left-right image of $U(1)_X$, where $U(1)_X'$ is broken by boundary conditions on the IR brane.


As discussed in the previous section, the third generation is a little different than the first two generations. In particular, the small bottom Yukawa coupling is explained by a hierarchy $\eta_{\bar{d}_3} / \eta_{\bar{u}_3} \sim 100$, with $\zeta_{\bar{u}_3}, \zeta_{\bar{d}_3} \sim 1$. Since $\zeta_{Q_3} \sim 1$ as well, all bulk multiplets are localized towards the IR brane, and generically (unlike in~\S\ref{sec:SUSYbreaking}) all third generation squarks are light, with masses generated by gaugino mediation and/or anomaly and radion mediation. 

The lepton sector can be constructed analogously to the quark sector, where the $\zeta_{L_i}$ are all of the same order to generate an anarchic PMNS matrix, with exponentially suppressed $\zeta_{\bar{\nu}_i}$ to realize light neutrino masses.

One issue with the above model is that the standard model Higgs fields are charged under $U(1)_X$, and spontaneously break it. Since $U(1)_X$ is also broken on the UV brane, this will lead to a problematic pseudo-Goldstone boson. (In CFT language, we have spontaneously broken an approximate flavor symmetry.) To avoid this issue, we replace $U(1)_X$ (and its left-right image $U(1)_X'$) with a discrete subgroup thereof, which is sufficient to forbid the problematic Yukawa couplings. In fact, for a different choice of charges, this discrete gauge symmetry could also explain the form of the NMSSM superpotential~(\ref{NMSSM1}). This has few physical consequences, however, so we do not comment on this possibility further.

We have so far ignored the possibility of intergenerational mixing in the UV brane mass terms/boundary conditions, (\ref{eqn:mixedBCs}), which could induce dangerous FCNCs. We will justify this assumption in~\S\ref{sec:flavor}, where we consider the inclusion of horizontal symmetries.

\subsection{Constraints from light exotics} \label{subsec:exoticCons}

While we have avoided the possibility (discussed in~\S\ref{subsec:unification}-\ref{subsec:SU5}) of weak-scale charged and/or colored exotics, the broken generators of $SU(2)_R\times U(1)_{B-L}$ will give rise to exotic gauge bosons with masses~(\ref{XYmass}) somewhat below the compactification scale. These exotic gauge bosons are observable at the LHC, and present LHC results already constrain the compactification scale.


The broken $SU(2)_R\times U(1)_{B-L}$ generators lead to light exotic gauge bosons $Z'$ and $W'_\pm$, with standard model quantum numbers $(\mathbf{1},\mathbf{1})_{0}$ and $(\mathbf{1},\mathbf{1})_{\pm 1}$. The $W'$ couples to right-handed fermions analogously to the way in which the standard model $W$ boson couples to left-handed fermions, whereas the $Z'$ couples to both left and right-handed fermions with couplings which depend on a possible IR-brane-localized kinetic mixing with the hypercharge generator. Kinetic mixing between the $W$ and $W'$ is also possible, but enters through higher dimensional operators suppressed by $(v/\Lambda_{\rm IR})^2 \lsim 10^{-4}$, and is therefore negligible. Despite the left-right symmetry, the gauge couplings for the $W'$ and $Z'$ may differ somewhat from their left-handed counterparts due to IR-brane localized kinetic terms.

There are a number of existing searches for $W'$ and $Z'$ gauge bosons at the LHC. The strongest limits on the $W'$~\cite{Wprime} and $Z'$~\cite{Zprime} gauge boson masses -- coming from leptonic decays $W' \to \ell \nu$ and $Z' \to \ell^+ \ell^-$ -- are above $3$ TeV, but these apply to the ``Sequential Standard Model''~\cite{Altarelli:1989ff}.
Constraints on the left-right model are typically somewhat weaker (see e.g.~\cite{Chatrchyan:2012it}, figure 6), and will depend on the IR-brane-localized kinetic terms as above. Moreover, the presence of superpartners such as the third generation squarks may dilute the branching fraction of the $W'$ and $Z'$ to leptons, whereas the presence of gauginos $\tilde{W}'$ and $\tilde{Z}'$ may further complicate the situation. A detailed phenomenological study would be required to establish the correct mass limits for our scenario, but we anticipate that masses below $2.5$--$3$ TeV will be ruled out, implying that the standard model KK modes lie in the $25$--$30$ TeV range, or higher. This implies a high cutoff and some degree of tuning from the quadratic divergence discussed in~\S\ref{sec:SUSYbreaking}.

Electroweak precision measurements also constrain the model.
However, a $Z'$ gauge boson mass of $2.5$--$3$ TeV is heavy enough to satisfy the bounds~\cite{EWPT}, hence these measurements give no new constraints.

\section{Flavor and horizontal symmetries} \label{sec:flavor}

So far we have relied on anarchic IR-brane Yukawa couplings and order-one bulk mass parameters to generate the observed Yukawa couplings and CKM matrix. However, this scenario can lead to dangerous flavor-changing neutral currents (FCNCs), e.g.\ via the exchange of KK gluons. In the mass basis, the off-diagonal KK gluon couplings are suppressed by $\zeta_i \zeta_j$ (see e.g.~\cite{Csaki:2008zd}), leading a suppression of FCNCs known as the ``RS-GIM mechanism.'' Nonetheless, applying the model-independent constraints of~\cite{bonatalk}, the authors of~\cite{Csaki:2008zd} find a constraint $m_{\rm KK} \gsim 21$ TeV in the usual non-supersymmetric RS model. We can apply the same constraints to the SUSY RS model with two caveats. Firstly, the experimental constraints on CP violation in the neutral kaon system have improved~\cite{bonatalk}, which we estimate to give a factor of two improvement, $m_{\rm KK} \gsim 40$ TeV, under the same assumptions as~\cite{Csaki:2008zd}. Secondly, the wavefunction profiles in the SUSY RS model are related to their ordinary RS counterparts via $\left[\zeta_Q \zeta_{\bar{u}}\right]_{\rm SUSY\ RS} = (\sin \beta)^{-1} \left[\zeta_Q \zeta_{\bar{u}}\right]_{\rm RS}$ and $\left[\zeta_Q \zeta_{\bar{d}}\right]_{\rm SUSY\ RS} = (\cos \beta)^{-1} \left[\zeta_Q \zeta_{\bar{d}}\right]_{\rm RS}$. Thus, the constraint is further enhanced by $(\cos \beta)^{-1}$, giving $m_{\rm KK} \gsim 60$ TeV for $\tan \beta =1$ and $m_{\rm KK} \gsim 140$ TeV for $\tan \beta = 3$.

Thus, KK gluon-mediated FCNCs provide a strong constraint on the compactification scale, leading to increased fine tuning. Moreover, the inclusion of additional bulk multiplets and an extended bulk gauge group studied in~\S\ref{sec:unification} introduces additional potential sources of FCNCs, including those mediated by the light exotic gauge bosons. In lieu of fully characterizing these effects, we instead look for a way to suppress the KK gluon FCNCs, in the hope that other sources of FCNCs will also be suppressed.

\subsection{Flavor alignment}

We consider mechanisms of partial flavor alignment, such as those explored in~\cite{flavor,Csaki:2008eh} in the non-supersymmetric context. In particular, we focus on the mechanism described in~\cite{Csaki:2008eh}, which admits a simple embedding in the left-right model considered above.

The basic idea is to align the down-type Yukawa couplings using a horizontal symmetry. A second horizontal symmetry can be used to align the bulk mass parameters, so that the primary sources of intergenerational mixing are the up-type Yukawa couplings. 
Since the dominant constraints on FCNCs come from the down-type sector, the constraint on $m_{\rm KK}$ is substantially relaxed.

In order to align the down-type sector without 
also aligning the up-type sector (which would eliminate CKM mixing)
it is necessary to introduce two quark doublets, $Q_u$ and $Q_d$, which couple to $\bar{u}$ and $\bar{d}$ respectively, where $Q_u$ is neutral under the horizontal symmetry and $Q_d$ carries a generation-dependent charge. To reproduce the standard model at low energies, the two quark doublets mix under off-diagonal UV-brane boundary conditions of the form~(\ref{eqn:mixedBCs}), which break the horizontal symmetry. To align the UV brane boundary conditions and the up-type bulk masses, a second horizontal symmetry -- broken on the IR brane -- is imposed in the bulk and on the UV brane.


\begin{table}[t]
\renewcommand{\arraystretch}{1.3}
\begin{center}
\begin{tabular}{c|cccccc}
& $Q_u$ & $Q_d$ & $U_{\bar{u}}$ & $U_{\bar{d}}$ & $H_u$ & $H_d$ \\
\hline
$U(1)_0$ & $\ell_i$ & $\ell_i$ & $r_i$ & $r_i$ & $\cdot$ & $\cdot$ \\
$U(1)_1$ & $p_i$ & $q_i$ & $-1-\tilde{p}_i$ & $1-\tilde{q}_i$ & $1$ & $-1$
\end{tabular}
\end{center}
\caption{Horizontal symmetries for flavor alignment, where $p_i$, $q_i$, $\tilde{p}_i$, $\tilde{q}_i$, $\ell_i$, $r_i$ are generation-dependent charges and $U(1)_0$ ($U(1)_1$) is broken on the IR (UV) brane. Choosing $\ell_i \ne \ell_j$ and $r_i \ne r_j$ for $i\ne j$ ensures that intergenerational mixing cannot occur in the bulk or on the UV brane.
\label{tab:horiz1}}
\renewcommand{\arraystretch}{1}
\end{table}
We now construct a model of this type
using the techniques of~\S\ref{subsec:split}, \S\ref{subsec:left-right}. As before, we start with a 3-2-2-1 model and later incorporate the left-right symmetry. We introduce two left-handed multiplets $Q_u$ and $Q_d$ as well as two right-handed multiplets $U_{\bar{u}} = (\bar{u},\bar{d}')$ and $U_{\bar{d}} = (\bar{u}',\bar{d})$, with the extra zero modes removed by mixed UV brane boundary conditions as in~(\ref{eqn:mixedBCs}). We impose two horizontal symmetries with the charge assignments shown in table~\ref{tab:horiz1}, where $U(1)_1$ assumes the role that $U(1)_X$ played in~\S\ref{subsec:PatiSalam}. We choose $\tilde{p}_i = p_i$ and $\tilde{q}_i = q_i$ to ensure that the Yukawa couplings
\begin{equation}
\begin{split}
\\[-2.5ex]
W_{\rm Yukawa} \subset (Q_u^1\bar{u}_1+Q_u^2\bar{u}_2+Q_u^3\bar{u}_3) H_u+(Q_d^1\bar{d}_1+Q_d^2\bar{d}_2+Q_d^3\bar{d}_3) H_d\,, \\[1ex]
\end{split}
\end{equation}
can be generated. Choosing $p_i$ and $q_i$ such that $q_i \ne q_j$ for $i \ne j$ and $p_i \notin \{p_j - 2, q_j, q_j-2\}$ for all $i,j$ ensures that no additional Yukawa couplings can appear in the down-type sector. The most general (quark-sector) Yukawa couplings allowed by the horizontal symmetries are then
\begin{equation}
\begin{split}
W_{\rm Yukawa} = (\hat{Y}_u)_i^j Q_u^i\bar{u}_j H_u+(\hat{Y}_u')_i^j Q_d^i\bar{u}_j' H_u + (\hat{Y}_d)_i^j Q_d^i\bar{d}_j H_d\,, \\[1ex]
\end{split}
\end{equation}
where $\hat{Y}_d$ is diagonal.

Since the bulk masses and UV brane boundary conditions are aligned as a consequence of $U(1)_0$, off-diagonal Yukawa couplings are needed in the up-type sector to generate a non-trivial CKM matrix. 
We first proceed in direct analogy with~\cite{Csaki:2008eh} by setting $p_i = p$ for all three generations, which allows anarchic $\hat{Y}_u$, where $\hat{Y}_u'$ can be forbidden by an appropriate choice of charges. The CKM matrix is then generated by $\hat{\zeta}_{Q_u^i}$, as in~\S\ref{sec:SUSYRS}, whereas the $\hat{\zeta}_{Q_d^i}$ are unfixed.

To realize a left-right embedding, we must unify the 3-2-2-1 bulk multiplets into left-right multiplets. The minimal approach is to combine $Q_d$ and $U_{\bar{d}}$ into a single multiplet, which sets $\hat{\zeta}_{Q_d^i} = \hat{\zeta}_{\bar{d}_i} \simeq \sqrt{y_d^i}$ without affecting the CKM matrix. We then introduce left-right partners $Q'$ for $U_{\bar{u}}$ and $U''=(\bar{u}'',\bar{d}'')$ for $Q_u$, choosing their $U(1)_1$ charges to forbid all Yukawa couplings to their components. For simplicity, we choose $r_i = - \ell_i$, so that $U(1)_0$ is left-right odd, forbidding the D-term deformation~(\ref{eqn:DtermDefm}). We add a left-right image $U(1)_1'$, broken by boundary conditions on the IR brane, and replace both $U(1)_1$ and $U(1)_1'$ with a discrete subgroup to avoid a pseudo-Goldstone boson.

The resulting model is similar to that of~\cite{Csaki:2008eh} with one important difference: in our case the right-handed up- and down-type quarks are mixed between the multiplets $U_{\bar{u}}$, $U_{\bar{d}}$ and $U''$ as required by 3-2-2-1 invariance of the UV-brane boundary conditions. This is problematic, however, as anarchic IR-brane-localized kinetic terms for $\bar{d}' \in U_{\bar{u}}$ can be generated, leading to off-diagonal KK gluon couplings $g_{d_R}^{i j} \propto \hat{\zeta}_{\bar{u}_i} \hat{\zeta}_{\bar{u}_j}$ in the right-handed down-type sector. If the coefficient of the IR-brane-localized kinetic term is order-one, then a very large KK gluon mass is needed to suppress the resulting FCNCs.\footnote{In~\cite{Csaki:2008eh} the IR-brane localized kinetic terms are assumed to be loop-suppressed, which relaxes the constraint on $m_{\rm KK}$ somewhat.}
 
The FCNC constraints come mainly from mixing between the first two generations ($K-\bar{K}$ mixing), with the weakest constraints on mixing between the second and third generations ($B_s-\bar{B}_s$ mixing). Thus, we can avoid this problem by choosing $p_1 \neq p_2 = p_3 \equiv p$, which prevents kinetic mixing involving the first generation. However, this sets $(\hat{Y}_u)^1_2 = (\hat{Y}_u)^2_1 = (\hat{Y}_u)^1_3 = (\hat{Y}_u)^3_1=0$, preventing CKM mixing with the first generation as well. We can compensate by choosing charge assignments such that certain off-diagonal elements of $\hat{Y}_u'$ are nonvanishing, but the effective couplings will then be $\mathcal{O}(\hat{\zeta}_{Q_d^i} \hat{\zeta}_{\bar{d}_j})$, which fails to reproduce the correct CKM mixing given the relation $\hat{\zeta}_{Q_d^i} = \hat{\zeta}_{\bar{d}_i} \simeq \sqrt{y_d^i}$ imposed above.

Instead, we take $p_1 \ne \tilde{p}_1$, which sets $(\hat{Y}_u)^1_1= 0$ but allows additional Yukawa couplings
\begin{equation}
\begin{split}
\\[-2.5ex]
W_{\rm Yukawa}' = W_{\rm Yukawa} + Q_u^1 \bar{u}_i' H_u + Q_d^i \bar{u}_1 H_u \,, \\[1.5ex]
\end{split}
\end{equation}
consistent with the absence of off-diagonal down-type Yukawa couplings and kinetic mixing involving the first generation.\footnote{These requirements are satisfied if $p \notin \{p_1,p_1-2,\tilde{p}_1,\tilde{p}_1+2\}$, $p_1 \ne \tilde{p}_1+2$, $q_i \notin \{p,p+2,p_1,\tilde{p}_1+2\}$ and $q_i \ne q_j$ for $i \ne j$. The couplings $\hat{Y}_u'$ are forbidden for $q_i - q_j \ne 2$ for all $i,j$.}
For example, we choose $p_1 = q_3 - 2$ and $\tilde{p}_1 = q_2$, so that
\begin{equation} \label{eqn:Yspecial}
\begin{split}
Y_u \sim
\begin{pmatrix}
0									& 0									& \hat{\zeta}_{Q^1_u} \hat{\zeta}_{\bar{d}_3} \\
\hat{\zeta}_{\bar{d}_2} \hat{\zeta}_{\bar{u}_1} 	& \hat{\zeta}_{Q^2_u} \hat{\zeta}_{\bar{u}_2}	& \hat{\zeta}_{Q^2_u} \hat{\zeta}_{\bar{u}_3} \\
0									& \hat{\zeta}_{Q^3_u} \hat{\zeta}_{\bar{u}_2}	& \hat{\zeta}_{Q^3_u} \hat{\zeta}_{\bar{u}_3} \end{pmatrix} \,,\;\;\;
Y_d \sim
\begin{pmatrix}
\hat{\zeta}_{\bar{d}_1}^2 	& 0						& 0 \\
0			 		& \hat{\zeta}_{\bar{d}_2}^2	& 0 \\
0					& 0						& \hat{\zeta}_{\bar{d}_3}^2 \end{pmatrix} \,, \\[1ex]
\end{split}
\end{equation}
up to order-one factors.
A simple choice of charges which ensures this structure is e.g. $p=-1$, $(q_1,q_2,q_3)=(3,0,-3)$. 

While the up-type Yukawa coupling matrix~(\ref{eqn:Yspecial}) contains several vanishing entries, this is consistent with the observed Yukawa couplings and CKM matrix in a special basis $Y_u = V_{\rm CKM}^{T} \mathrm{diag}(y_u, y_c, y_t) V_{\bar{u}}^{\dag}$, $Y_d = \mathrm{diag}(y_d,y_s,y_b)$, where $V_{\bar{u}}$ is chosen to set these entries to zero. In particular, employing the Wolfenstein parameterization~\cite{Wolfenstein:1983yz} and working to leading order in $\lambda$, $m_c/(\lambda^2 m_t)$, and $m_u/(\lambda^3 m_t)\ll 1$, we find (up to a choice of unphysical phases)
\begin{equation} \label{eqn:Yspecial1}
\begin{split}
Y_u \simeq \begin{pmatrix}
0 & 0 & A \lambda^3 (1-\rho - i \eta) y_t \\
\frac{1}{\lambda}y_u & -\frac{\rho + i \eta}{1-\rho-i \eta} y_c & -A \lambda^2 y_t \\
0&\frac{1}{A \lambda^2 (1-\rho-i \eta)}y_c&y_t\end{pmatrix}
\sim
\begin{pmatrix}
0 & 0 & \lambda^3 y_t \\
y_u/\lambda & y_c & \lambda^2 y_t \\
0 & y_c/\lambda^2 & y_t
\end{pmatrix} \,, \\[1ex]
\end{split}
\end{equation}
where $V_{\bar{u}}$ (which can be computed explicitly) is similar to~(\ref{diagmatrix}).
Thus, we can reproduce the correct Yukawa couplings and CKM matrix for (cf.~(\ref{wffactor}))
\begin{equation}
\begin{split}
\\[-2.5ex]
\begin{array}{lll}
\hat{\zeta}_{Q^1_u}  \simeq \sqrt{\frac{v \cos \beta}{m_b}}\frac{\lambda^3 m_t}{v \sin \beta}, &	\hat{\zeta}_{Q^2_u} \simeq \lambda^2 \hat{\zeta}_{Q^3_u},	 &  \\[2.5ex]
\hat{\zeta}_{\bar{u}_1} \simeq \sqrt{\frac{v \cos \beta}{m_s}} \frac{m_u}{\lambda v \sin \beta}, & \hat{\zeta}_{\bar{u}_2} \simeq \frac{m_c}{\lambda^2 \hat{\zeta}_{Q^3_u} v \sin \beta},
& \hat{\zeta}_{\bar{u}_3} \simeq \frac{m_t}{\hat{\zeta}_{Q^3_u} v \sin \beta}, \\[2.5ex]
\hat{\zeta}_{\bar{d}_1} \simeq \sqrt{\frac{m_d}{v \cos \beta}}, & \hat{\zeta}_{\bar{d}_2} \simeq \sqrt{\frac{m_s}{v \cos \beta}},
& \hat{\zeta}_{\bar{d}_3} \simeq \sqrt{\frac{m_b}{v \cos \beta}}, \\[1.5ex]
\end{array}
\end{split}
\end{equation}
where $\hat{\zeta}_{Q^3_u}$ is a free parameter.

By construction, kinetic mixing on the IR brane is only allowed between the second and third generations, and the dominant FCNC constraints will come from the up-type sector, as in~\cite{Csaki:2008eh}. Foregoing a detailed analysis along the lines of~\cite{Csaki:2008zd,Csaki:2008eh}, we estimate a bound $m_{\rm KK} \gsim 10$ TeV by analogy with~\cite{Csaki:2008eh}, allowing some leeway for improved experimental bounds~\cite{bonatalk} and somewhat different bulk profiles in our case. This constraint is subdominant to the constraints from the non-observation of exotic gauge bosons considered in~\S\ref{subsec:exoticCons}.

\subsection{R-parity violation revisited}

We comment briefly on the status of R-parity violation in models of this type. Since $U(1)_{B-L}$ is gauged on the UV brane, R-parity violation is forbidden there (c.f.~\S\ref{subsec:BNV}). Moreover, many dangerous proton decay operators can potentially be forbidden on either brane by an appropriate choice of the charges for $U(1)_0$ and $U(1)_1$. This raises the question (which we defer to a future work) of whether proton decay can be forbidden without imposing $\mathbb{Z}_3^{(L)}$, allowing Majorana neutrino masses.

R-parity violation can still occur on the IR brane, but the form of the R-parity violating couplings is constrained by $U(1)_1$. In particular, not all of the couplings in table~\ref{RPVnumerical} will be generated, with the available couplings depending on the choice of $p_i$, $\tilde{p}_i$ and $q_i$. Combined with the presence of $\tilde{b}_R$ in the effective theory due to the mechanism of~\S\ref{subsec:split}, the R-parity violating phenomenology may be substantially altered relative to the discussion of~\S\ref{sec:RPV}, offering at the same time model-building flexibility and the potential for dangerous operators. We defer a complete consideration to a future work.

\section{Conclusions} \label{sec:conclusions}

In this paper, we have considered the possibility of a natural supersymmetric model where the Higgs boson is protected from the effects of a relatively heavy gluino by compositeness. Only a few superpartners need appear below the confinement scale to solve the little hierarchy problem, whereas the large hierarchy is explained by compositeness, and the remaining superpartners can safely decouple without introducing fine tuning. This approach therefore provides a simple realization of the ``natural SUSY'' paradigm -- where the Higgsinos, gluino and stops are light and the first and second generation squarks are heavy -- allowing a natural model consistent with present LHC searches.

For definiteness, we considered a supersymmetric Randall-Sundrum model, related to a four-dimensional composite model by the AdS/CFT correspondence. Motivated by the RS flavor problem, we placed the standard model fermions and gauge bosons in the bulk, with the Higgs on the IR brane and the Yukawa couplings explained by the exponential profiles of the bulk fermions. This gives rise to the RS GIM mechanism and provides a partial flavor protection. To protect the Higgs sector from large splittings -- as demanded by naturalness -- 
 we assumed that supersymmetry is broken dynamically on the UV brane. (The dual picture is that of a SUSY-breaking sector weakly coupled to a supersymmetric confining theory, introducing small splittings into the composite states.)

With these assumptions, we have argued that the first and second generation squarks are generically decoupled due to their proximity to the SUSY breaking sector. This conclusion is somewhat model dependent, but violating it typically requires introducing additional independent scales into the problem to ensure that the stop is not degenerate with the other squarks (to evade strong constraints on the first two generations) without simply decoupling them. Hence, we are led almost inevitably to a natural SUSY spectrum, a scenario we refer to as ``warped natural SUSY.''

The absence of the first two generations of squarks from the low-energy effective theory in models of warped natural SUSY leads to a sizable two-loop $g_3^4$ quadratic divergence in the stop mass, correcting the Higgs mass at three loops. This correction competes with a similarly-sized two-loop $g_2^4$ quadratic divergence in the Higgs mass, and there is a partial cancellation between the two terms. Nonetheless, fine tuning increases rapidly as the cutoff is raised above $10$ -- $15$ TeV, beyond which the quadratic divergence becomes the dominant source of tuning.

The large splittings in the elementary fields can induce dangerous radiative corrections to the hypercharge $D$-term, which is a gauge-singlet. To avoid this catastrophe, we are forced either to extend the standard model gauge group to a semi-simple group, or to include an outer automorphism under which all $U(1)$ factors are charged (which also requires the connected component of the gauge group to be extended). To ensure cancellation of the radiative corrections down to the weak scale, we assume that the extended gauge group is only broken on the IR brane, i.e. by the confining dynamics.

Due to the extended gauge symmetry, the standard model fermions must be embedded into larger multiplets. This ``unification'' carries two hazards. It can lead to proton decay or other rare processes in some instances where two or more standard-model fermions occupy the same multiplet. Moreover, the introduction of new charged and/or colored particles to fill out the multiplets can lead to light exotic fermions, in contradiction with LEP results. The splitting between different standard model representations in the UV-brane localized (fundamental) multiplets is exponentially suppressed, and we argue that only neutral, colorless exotics are permissible. This immediately rules out $SU(5)$-based models, which require split multiplets to avoid excessive proton decay, as well as the Pati-Salam model, which requires split multiplets (or a high Pati-Salam breaking scale) to avoid excessive rare meson decays.

Based on an informal classification, we conclude that only the minimal left-right model with a gauged left-right symmetry (and groups containing it) can solve the problem without introducing light exotics. Even so, the $W'$ and $Z'$ gauge bosons from left-right breaking will appear well below the confinement scale, and consequently present LHC searches for these gauge bosons place an indirect lower bound on the confinement scale of roughly $30$ TeV. This implies fine tuning of the Higgs potential of the order of $5\%$ from the two-loop quadratic divergence discussed above.

We have constructed an example left-right model as a proof of principle that the correct Yukawa couplings can be reproduced in this framework. While the left-right multiplets remain unified, their Yukawa couplings are split by introducing two bulk multiplets with different profiles and off-diagonal boundary conditions on the UV brane along with an additional $U(1)$ symmetry (broken on the UV brane) which controls which multiplets can coupled to the up- and down-type Higgs on the IR brane. As in the usual non-supersymmetric RS model, more work is required to adequately suppress FCNCs with a low confinement scale. As a further proof of principle, we have constructed an explicit model based on horizontal symmetries which can be naturally incorporated into the above scenario, and which ensures that FCNCs are sufficiently suppressed.  

Another generic problem with composite models is the potential to generate large dimension-six proton decay operators in the low energy effective theory, suppressed only by the confinement scale. These operators are R-parity even, and hence some further symmetry is needed to prevent their appearance. In this work we have imposed a simple $\mathbb{Z}_3^{(L)}$ for definiteness, which requires Dirac neutrino masses. In the presence of this symmetry, R-parity is no longer required to prevent proton decay, and it is natural to consider R-parity violation. We have shown that the assumption of bulk fermions and anarchic couplings on the IR brane naturally leads to a model with sufficient R-parity violation to avoid missing energy or displaced vertices at the LHC, but with small enough baryon number violation to satisfy bounds on dinucleon decay and $n$-$\bar{n}$ oscillations. 

The introduction of R-parity violation relaxes LHC constraints on the stop mass, while removing the WIMP dark matter candidate common to R-parity conserving models. However, as discussed in~\S\ref{sec:Higgs}, we find that the DFSZ axion model can be naturally incorporated into the SUSY RS model, solving the strong CP problem and providing dark matter candidates in the axion and/or the axino. This possibility deserves more study.

We conclude that a supersymmetric composite model of the kind considered in this work is a viable model of natural supersymmetry. While the necessity of extending the gauge group combined with the present LHC constraints on exotic gauge bosons leads to some tuning, this tuning is relatively mild, and it is possible that violating one of our assumptions could lead to a completely natural model.

A number of important questions remain to be explored. We have not fully addressed the coincidence problem between the confinement scale and the gaugino masses, and a natural mechanism explaining this coincidence would be a boon to supersymmetric confining theories in general. Moreover, our analysis relies on several assumptions, e.g. about how the RS flavor problem is solved, and it would be interesting to understand whether different assumptions would lead to similar conclusions or not. Our mechanism for suppressing proton decay requires Dirac neutrino masses, but other options exist. For instance, the horizontal symmetries we impose to fully solve the flavor problem could play a role in suppressing proton decay, analogous to~\cite{MFVSUSY}.

The cosmological implications of models of this type remain to be explored.  
For instance, baryonic RPV couplings may wash out any primordial baryon density and require baryogenesis below the electroweak scale.
This kind of low scale baryogenesis was discussed in~\cite{baryogenesis}, and it would be interesting to embed this mechanism into the framework of warped natural SUSY. 
Furthermore, the confining phase transition may occur too slowly, leading to an underpopulated universe~\cite{Creminelli:2001th}, and requiring very low scale inflation with a reheating temperature below the phase transition. 
 However, workarounds may exist which allow high-scale inflation~\cite{SundrumPC}.

\section*{Acknowledgements}
We thank Matthew Reece and Lisa Randall for extended discussions and comments on the manuscript. BH is supported by the Fundamental Laws Initiative of the Harvard Center for the Fundamental Laws of Nature.
YN is supported by JSPS Fellowships for Young Scientists.

\end{document}